\documentclass[reprint,showpacs,preprintnumbers,amsmath,amssymb]{revtex4-1}

\pdfoutput=1
\usepackage[pdftex]{graphicx}
\usepackage[unicode=true, bookmarks=false,bookmarksnumbered=true,bookmarksopen=true, breaklinks=false,backref=false,pagebackref=false, colorlinks=true]{hyperref}
\hypersetup{pdftitle={ },pdfauthor={ },linkcolor=black,citecolor=black, urlcolor=black, filecolor=black,pdfpagelayout=OneColumn,pdfnewwindow=true,pdfstartview=XYZ, plainpages=false}
\usepackage{color}

\usepackage{braket}
\usepackage[T1]{fontenc}
\usepackage{amsmath}
\usepackage{amsthm}
\usepackage{amssymb}
\usepackage{url}
\usepackage[english]{babel}
\usepackage{dcolumn}
\usepackage{bm}
\usepackage{subfig}
\usepackage{float}
\usepackage{url} 
\usepackage{grffile}

\newcommand{\Tr}{\mbox{\rm Tr\,}}

\newcommand\T{\rule{0pt}{2.6ex}}
\newcommand\B{\rule[-1.2ex]{0pt}{0pt}}

\usepackage{cleveref}
\crefname{section}{Section}{Sections}
\Crefname{section}{Section}{Sections}
\crefname{equation}{Eq.}{Eqs.}
\Crefname{equation}{Eq.}{Eqs.}
\crefname{figure}{Fig.}{Figs.}
\Crefname{figure}{Fig.}{Figs.}
\crefname{table}{Table}{Tables}
\Crefname{table}{Table}{Tables}

\graphicspath{{lastfig/},{figures/},{}}
\usepackage{verbatim}

\usepackage{graphicx}
\begin{document}

\title{Pure gauge QCD Flux Tubes and their widths at Finite Temperature}
\author{P. Bicudo}
\email{bicudo@tecnico.ulisboa.pt}
\author{N. Cardoso}
\email{nuno.cardoso@tecnico.ulisboa.pt}
\author{M. Cardoso}
\email{mjdcc@cftp.ist.utl.pt}
\affiliation{CFTP, Departamento de F\'{\i}sica, Instituto Superior T\'{e}cnico
(Universidade T\'{e}cnica de Lisboa),
Av. Rovisco Pais, 1049-001 Lisboa, Portugal}

\begin{abstract}
We study the flux tubes produced by static quark-antiquark,
quark-quark and quark-adjoint charges at finite temperature in pure gauge SU(3) lattice QCD.
This is relevant both for the study of flux tubes and strings, and for the interaction of heavy quarks and other colour sources in heavy ion collision physics. 
Our sources are static and our lattice correlators are composed of fundamental and adjoint Polyakov loops.
To signal the flux tubes, we compute the square densities of the chromomagnetic and chromoelectric fields with plaquettes, in a gauge invariant framework.
We study the existence and non-existence of flux tubes both below and above the deconfinement phase transition temperature $T_c$.
Using the Lagrangian density as a profile distribution, we also compute the widths of the flux tubes and study their widening as a function of the inter-charge distance. We determine our results with both statistical and systematic errors.

\end{abstract}
\maketitle

\section{Introduction}


The understanding of confinement  and deconfinement in QCD remains a central problem of particle physics. A major evidence of QCD confinement is the flux tube arising
between quark-antiquark static charges, both from gauge invariant pure gauge lattice QCD simulations \cite{DiGiacomo:1989yp,DiGiacomo:1990hc,Singh:1993jj,Bali:1994de}
 and from experimental observations like Regge trajectories
 \cite{Regge:1959mz,Regge:1960zc,Collins:1977jy,Kaidalov:2001db,Bugg:2004xu} consistent with linear confining potentials.
Even in dynamical QCD where the flux tube breaks due to the creation of another quark and antiquark, a flux tube develops up to moderate quark-antiquark distances. 
Different flux tubes have also been shown to occur in lattice QCD simulations of different exotic hadrons, typical of $SU(3)$ 
\cite{Cardoso:2009kz,Cardoso:2011fq,Cardoso:2011cs,Cardoso:2012rb}. 
It is important for the fundamental understanding of the pure gauge QCD flux tubes to measure the flux tube profile \cite{Polyakov:1975rs,Banks:1977cc,Smit:1989vg,Bali:1996dm,Gubarev:1999yp,Koma:2003gq,Chernodub:2005gz} with more quantitative results.

Moreover, in the topic of High Energy Heavy Ion Physics, the interactions between heavy quarks are relevant for the hard probes of the QCD phase diagram \cite{Matsui:1986dk}, and the development of flux tubes between different charges may help to microscopically understand the phenomenological vortex line model for the flow and fragmentation  \cite{Andersson:1986gw}. Thus it is also important to study flux tubes, not only at zero temperature, but also at finite temperature.

In this work, we study whether flux tubes exist or not, between different static charges,  at different temperatures above and below pure gauge QCD critical deconfinement temperature $T_c$.
In particular, we  study the flux tubes created between static quark-antiquark, quark-quark and quark-adjoint  charges. 

Notice we do not know exactly, neither the theoretical origin of the QCD flux tubes not their effective behaviour, and it is important to explore in more detail their properties in order to, hopefully, one day solve this important problem. 
For instance, two qualitatively different effective models for the QCD flux tube exist.

\begin{figure}[!t]
\centering
\includegraphics[width=1.0\columnwidth]{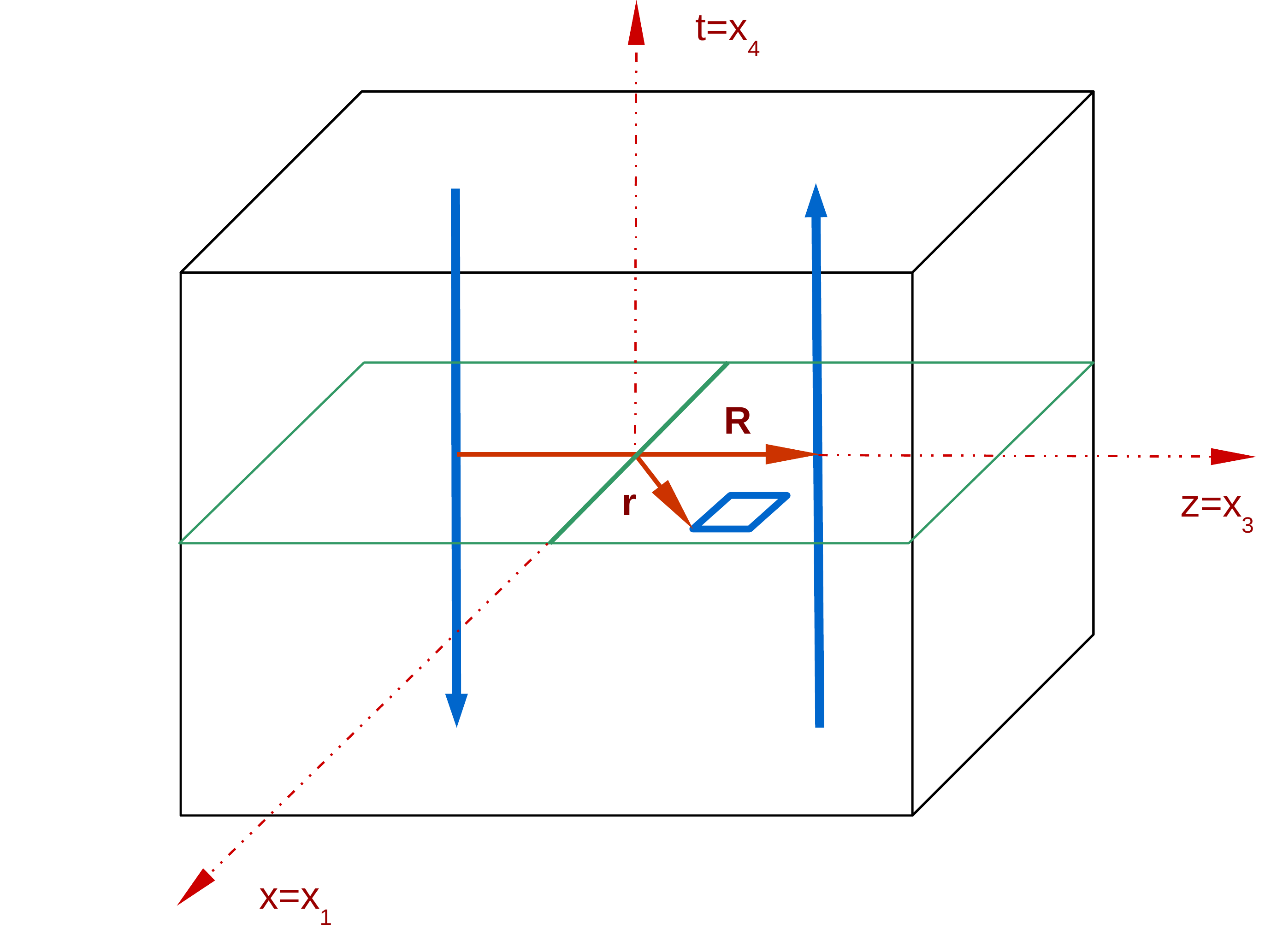}
\caption{ 
\color{black}
(Colour Online.) Geometry of the Polyakov loops and plaquette illustrated in the $Q \bar Q$ case. 
The lattice is represented in black, the Polyakov loops and plaquette are painted in blue and the axis and relevant vectors are painted in red. $\mathbf R$ is the vector separating the two Polyakov loops,and $\mathbf r$ is the position of the plaquette. Because the Euclidian space-time is four-dimensional, we represent only one temporal dimension and two spatial dimensions. The plane in green is the projection of the three-dimensional space into the $x,z$ dimensions. The charge axis is the $z$ axis. The mediator plane is the $x,y$ plane, here represented by a green line only since we do not show the $y$ dimension.}
\label{fig:geometryFiniteT}
\end{figure}

Already in the 1970's,  Nambu \cite{Nambu:1974zg}, 't Hooft \cite{'tHooft:1979uj} and Mandelstam \cite{Mandelstam:1974pi}
proposed that quark confinement would be physically interpreted using the dual version of the superconductivity
\cite{Baker:1989qp,Baker:1991bc}. The QCD vacuum state would behave like an ordinary superconductor,
where Cooper-pair condensation leads to the Meissner effect, and the magnetic flux is excluded from the vacuum and squeezed in a
quasi-one-dimensional tube, the Abrikosov-Nielsen-Olesen vortex, where the magnetic flux is quantized topologically \cite{Abrikosov:1956sx,Nielsen:1973cs,Cardoso:2006mf}.
In a superconductor, the fields are approximately classical and the flux tube main parameter is the penetration length $\lambda$ in the London equation has a direct relation with an effective mass of the interaction particle fields,
i. e., the photon.
In QCD, the dual gluon mass has been studied by several authors, \cite{Burdanov:1998tf,Jia:2005sp,Suzuki:2004uz,Suganuma:2004gq,
Suganuma:2004ij,Suganuma:2003ds,Kumar:2004fj,Burdanov:2002ne}, as well as the gluon effective mass  \cite{Cardoso:2010kw}.
Interestingly, there is also an evidence for a gluon mass in the Landau Gauge 
\cite{Oliveira:2010xc} 
and in the multiplicity of particles produced in heavy ion collisions
\cite{Bicudo:2012wt}. 
Recently,  the penetration length started to be computed with gauge invariant lattice QCD techniques 
\cite{Cardaci:2010tb,Cea:2012qw,Cardoso:2013lla}.


On the other extreme limit, at quark-antiquark distances larger than the penetration length, the flux tube is similar to a quantum string, contrary to the picture of the superconductor with classical electromagnetic fields.
Due to its vibration, the quantum string has a Gaussian profile and a finite width, different from the exponential profile of the classical superconductor flux tube
\cite{Luscher:1980iy,Munster:1980me}.
Thus a second model of the QCD flux tube is given by the
string model, based on the \foreignlanguage{american}{Nambu-Goto} Action
\cite{Nambu:1978bd,Goto:1971ce}.

At zero temperature, the energy of the quantum string with length $R$ and fixed ends, with quantum transverse fluctuations quantum number $n$, is
expressed in the L\"uscher term  and in the Arvis Potential \cite{Luscher:1980ac,Arvis:1983fp},
\begin{eqnarray}
V_{n}(R)&=&\sigma\sqrt{R^{2}+\frac{2\pi}{\sigma}(n-\frac{D-2}{24})}
\nonumber \\
&=&\sigma R+\frac{\pi}{R}(n-\frac{D-2}{24})+\ldots
\label{Arvis}
\end{eqnarray}
indeed observed in lattice QCD for 4 space-time dimensions  \cite{Luscher:1980iy}.
In Eq. (\ref{Arvis}), $D$ is the dimension of the space time. Note that the Arvis potential is tachyonic at small distances since
the argument of the square root becomes negative in this limit, moreover rotational invariance is only achieved for $D=26$.
Nevertheless,  the first two terms in the $1/R$ expansion, $\sigma R + \pi (n-\frac{D-2}{24}) {1 \over R}$, are more general than the Arvis potential, since they fit the $D=3$ and
$D=4$ lattice data quite well down to distances much smaller than the Arvis tachyonic distance.
The $1/R$ term is independent of the string tension $\sigma$ and for the physical
$D=3+1$ has the value $-\frac{\pi}{12}$. This is the L\"uscher term  \cite{Luscher:1980ac}.
The energy spectrum of a static quark-antiquark and of its flux tube is certainly well defined (not tachyonic) and this was
the first evidence of flux tube vibrations found in lattice field theory. Moreover it was shown \cite{Luscher:1980iy} that
the width of the ground state  flux tube  diverges when $R\rightarrow\infty$ with a logarithmic law,
\begin{equation}
w^{2}\sim w_{0}^{2}\,\log\frac{R}{R_{0}}
\end{equation}
where the squared width $w^{2}$ is the mean squared radius of the flux tube, computed in its centre. 
This enhancement of the flux tube transverse radius as $R\rightarrow\infty$ is frequently called
{\em widening}. The widening as been recently extended with two-loop calculations \cite{Gliozzi:2010zt}.
The flux tube widening has been verified numerically
for compact U(1) QED $D=3+1$ lattices \cite{Amado:2012wt}, for non-abelian SU(2) $D=2+1$ lattices
\cite{Armoni:2008sy,Armoni:2008sy,Bornyakov:2002vt,Bornyakov:2003gn,Cardoso:2012aj,Caselle:1995fh,
Caselle:2010zs,Caselle:2012rp,Chernodub:2007wi,Giudice:2006xe,Giudice:2006hw,Gliozzi:2006sj,Gliozzi:1994bc,Gliozzi:2010jh,
Gliozzi:2010zv,Greensite:2000cs,Lucini:2001nv,Meyer:2010tw} and, more recently, for the more physical four-dimensional
$SU(3)$ case \cite{Bakry:2010zt,Bakry:2010sp}.
Recently, it has been shown the flux tubes exhibit characteristics of both superconductor and string models, with both a penetration length $\lambda$ and the quantum widening of $w$ 
\cite{Cardoso:2013lla}.


Moreover, at finite $T$, close to $T_c$ but still in the confining regime of $T < T_c$, it has been predicted in Ref. \cite{Allais:2008bk} that widening becomes linear with the inter-charge distance R. This occurs because the flux tube width $w$ also depends on the extent $\tau$ of the compactified time distance of the lattice,
\begin{equation}
\sigma w^2 = {1 \over 2 \pi} \log { \tau \over \tau_c} + { R \over 4 \tau} - {1 \over \pi} e^{- 2 \pi {R \over \tau}}   + \cdots
\label{eq:linearfiniteT}
\end{equation} 
Since $\tau$ is small at finite $T$, we expect the dominant term to be the linear term in $R$.
This result has been verified for the Ising model  \cite{Allais:2008bk} and for compact $U(1)$ \cite{Amado:2012wt}. Recently, widening for $SU(3)$ has also been studied for Baryons \cite{Bakry:2010zt,Bakry:2014gea,Bakry:2015csa}
We  intend to test the linear broadening of \cref{eq:linearfiniteT} as well for $SU(3)$.

However, at $T>T_c$, it has been recently claimed by Refs. \cite{Shibata:2015ywl,Cea:2015wjd} that a flux tube continues to exist. This apparently contradicts previous results on the static quark-antiquark $Q  \bar Q$  potentials which indicate that linear confinement disappears for $T > T_c$ \cite{Kaczmarek:1999mm}. Thus, we also intend to clarify how the colour fields and possible flux tubes behave above the phase transition.


\begin{table}[!t]
\begin{ruledtabular}
\begin{tabular}{cc|ccc}
\T\B Volume & $\beta$ & $T/T_c$	&	$a\sqrt{\sigma}^\text{\cite{Edwards:1997xf}}$	&	\# config.	\\ \hline
\hline
\T\B $32^4$ & 6.0 & 0	&	0.219718	&	1100	\\ \hline
\T\B $48^3\times8$ & 5.96 & 0.845	&	0.235023	&	5990	\\ \hline
\T\B $48^3\times8$ & 6.0534 & 0.986	&	0.201444	&	5990/5110* 	\\ \hline
\T\B $48^3\times8$ & 6.13931 & 1.127	&	0.176266	&	5990 	\\ \hline
\T\B $48^3\times8$ & 6.29225 & 1.408	&	0.141013	&	5990 	\\ \hline
\T\B $48^3\times8$ & 6.4249 & 1.690	&	0.117513	&	5990 	\\ 
\end{tabular}
\end{ruledtabular}
\caption{Lattice ensembles, in $48^3\times 8$ volumes at finite $T$ and in a $32^4$ volume at $T=0$.
We denote with an $^*$ the number of remaining configurations after we remove the configurations contaminated by the other phase.
}
\label{tab:latticesimdata}
\end{table}

Here we extend our previous study \cite{Cardoso:2013lla} to finite temperature $T$. We aim 
to measure in detail with lattice QCD techniques the profile of SU(3) pure gauge flux tubes in dimensions $D=3+1$.
We study the colour fields distributions inside the flux tubes formed by Polyakov loops in the static $Q\bar{Q}$, $QQ$ and $QA$ 
\footnote{
\textcolor{black}{ In our notation, $A$ stands for adjoint SU(3) representation charge, the same colour charge of a gluon, whereas the quark has a fundamental representation colour charge.}
}
systems at finite $T$, both below and above the phase transition temperature $T_c$.
We address how the flux tube evolves with the distance between quarks and when the temperature increases beyond the phase transition.
Moreover, we compare our results with the $T=0$ flux tubes with the static $Q\bar{Q}$ system computed with the Wilson loop \cite{Cardoso:2013lla}.

In Section II, we describe the lattice formulation at finite $T$. We briefly review the Polyakov loops of the different colour charges systems, detail how to compute the colour field and Lagrangian distributions, and discuss the techniques we utilize to increase the signal over noise ratio. 
In Section III, we show our results for the different squared field densities, both in the charge axis and in the mediator plane,  and qualitatively discuss them.
In Section IV,  we compute and analyse the widening of the $Q\bar{Q}$ flux tube profile in the inter-charge mediator plane when the separation of the charges increase, in particular we analyse the systematic errors of the width and combine them with the statistical errors. 
%
Finally, we present our conclusion in Section V.

\section{$SU(3)$ lattice QCD framework}

{\color{black}
We aim to measure the $SU(3)$ colour flux tube produced by static charges, in a lattice QCD framework. 
We utilize the quenched QCD configurations detailed in Table \ref{tab:latticesimdata}.

Our two charges are separated by a lattice vector (four-dimensional) $\mathbf R$ with spacial components only. We choose our frame with the charge axis in the $z$ direction and the median point of the charges as the origin. The mediator plane is the $x,y$ plane. Moreover we have the Euclidian time axis in the fourth dimension $t$. This is illustrated in Fig. \ref{fig:geometryFiniteT}.

The relevant observables of the flux-tube system can be extracted from the correlation of the plaquette
$\square_{\mu\nu}$ and charge operators $\cal O$. The plaquette measures the fields and is computed with four gauge links $U$,
\begin{equation}
\square_{\mu\nu}(\mathbf r) = \frac{1}{3} \Tr\left[U_\mu(\mathbf r)U_\nu((\mathbf r+\hat e_\mu)U^\dagger_\mu(\mathbf r+ \hat e_\nu)U^\dagger_\nu(\mathbf r)\right] \ .
\end{equation}
where $\mathbf r$ is the four-dimensional position of the plaquette, see Fig. \ref{fig:geometryFiniteT}.

We aim to compare the field density inside the flux tube to the field density in the vacuum. For the vacuum we utilize a reference point $\mathbf r_\text{ref}$ sufficiently far from the centre of the flux tube.
We measure the following correlator \cite{Peng:1993ip},
\begin{equation}
f_{\mu\nu}(\mathbf R,\mathbf r) = \frac{\beta}{a^4} \left[\frac{\Braket{\mathcal{O(\mathbf R)}\,\square_{\mu\nu}(\mathbf r)}-\Braket{\mathcal{O(\mathbf R)}\,\square_{\mu\nu}(\mathbf r_\text{ref})}}{\Braket{\mathcal{O(\mathbf R)}}}\right] \ ,
\label{eq:fmunucomp}
\end{equation}
}

Our operators $\cal O$ are combinations of fundamental representation Polyakov loops $L$,
\begin{eqnarray}
\mathcal{O}=& \ L^\dagger(- \mathbf R / 2)\,L(\mathbf R / 2)\quad\quad & \text{for the } Q\bar{Q} \text{ system} \ ,
\nonumber
\\
\nonumber
\mathcal{O}=& \ L(- \mathbf R / 2)\,L(\mathbf R / 2)\quad\quad \ & \text{for the } QQ \text{ system} \ , 
\\
\nonumber
\mathcal{O}=&\biggl[ L(- \mathbf R / 2)L^\dagger(- \mathbf R / 2)& \hspace{-3pt} -1 \biggr]\, L(\mathbf R / 2)
\\
&& \text{for the } QA \text{ system} \ ,
\label{eq:polyakov_loops}
\end{eqnarray}
where  $A$ stands for a static charge in the adjoint representation of $SU(3)$, and
\begin{equation}
L(\mathbf R/2)=\frac{1}{3}\Tr\Pi_{t=1}^{N_t}U_4(\vec R/2,t)
\end{equation}
is the fundamental Polyakov loop 
and $N_t$ is the number of time slices of the lattice. 

We also use the periodicity in the time direction for the Polyakov loops
 to average the plaquette over the time direction,
\begin{equation}
\square_{\mu\nu}(\mathbf r) = \frac{1}{N_t} \sum_{t=1}^{N_t} \square_{\mu\nu}(\vec r,t) \ .
\end{equation}

\begin{figure}[!t]
\centering
\includegraphics[width=1.0\columnwidth]{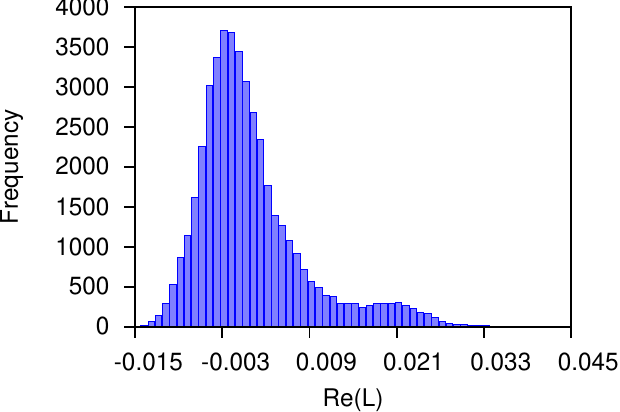}
\caption{ (Colour Online.) Histogram of the Polyakov loop history for $\beta=6.055$.}
\label{fig:histpolyloop}
\end{figure}

\begin{figure*}[!t]
\captionsetup[subfloat]{farskip=0.1pt,captionskip=0.1pt}
\begin{centering}
\subfloat[squared densities in the charge axis at $T=0.845\,T_c$.\label{fig:Field_XY_5p96}]{
\begin{centering}
\includegraphics[width=8cm]{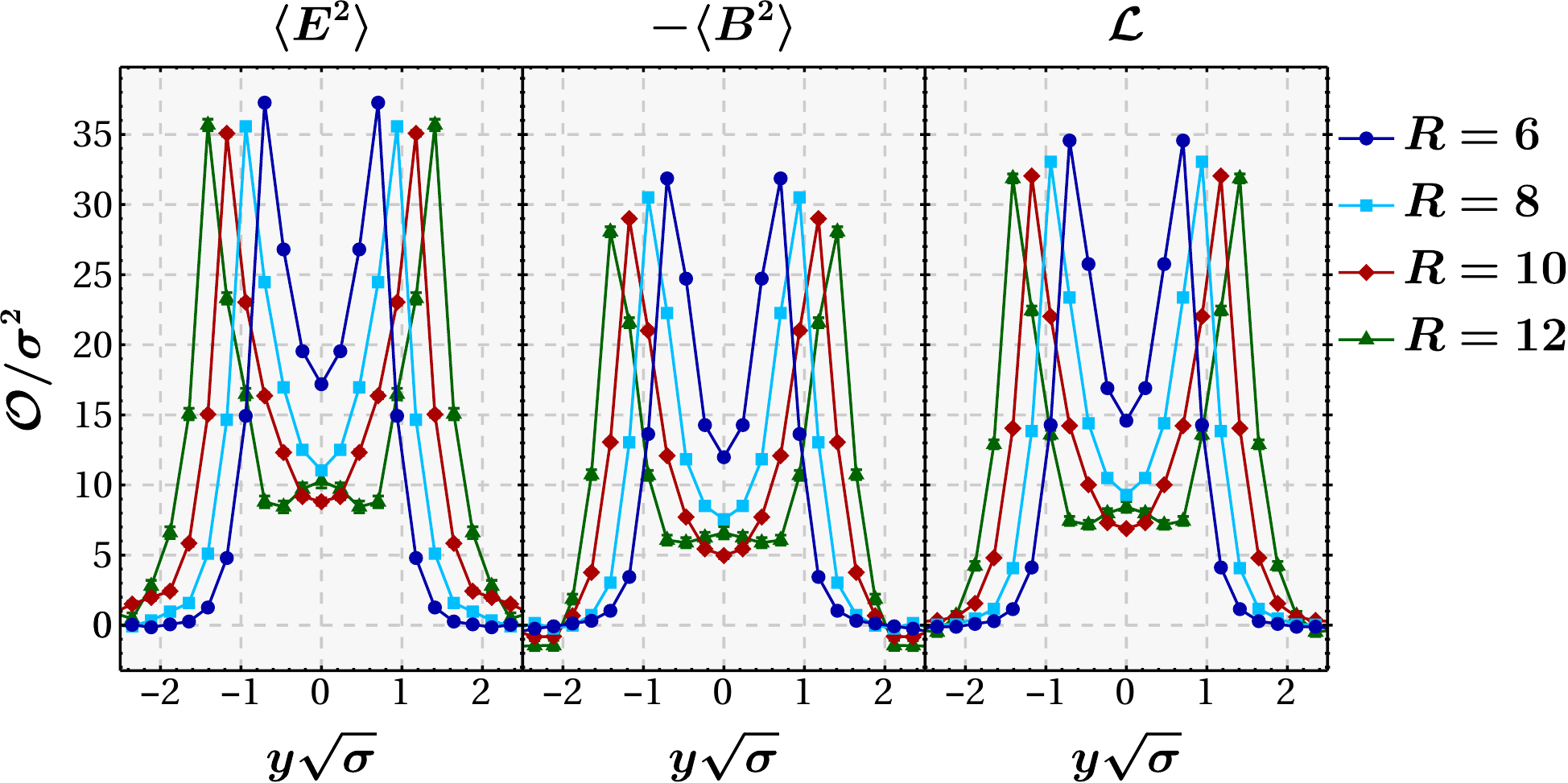}
\par\end{centering}}
\subfloat[squared densities in the mediator plane at $T=0.845\,T_c$.\label{fig:Field_XZ_5p96}]{
\begin{centering}
\includegraphics[width=8cm]{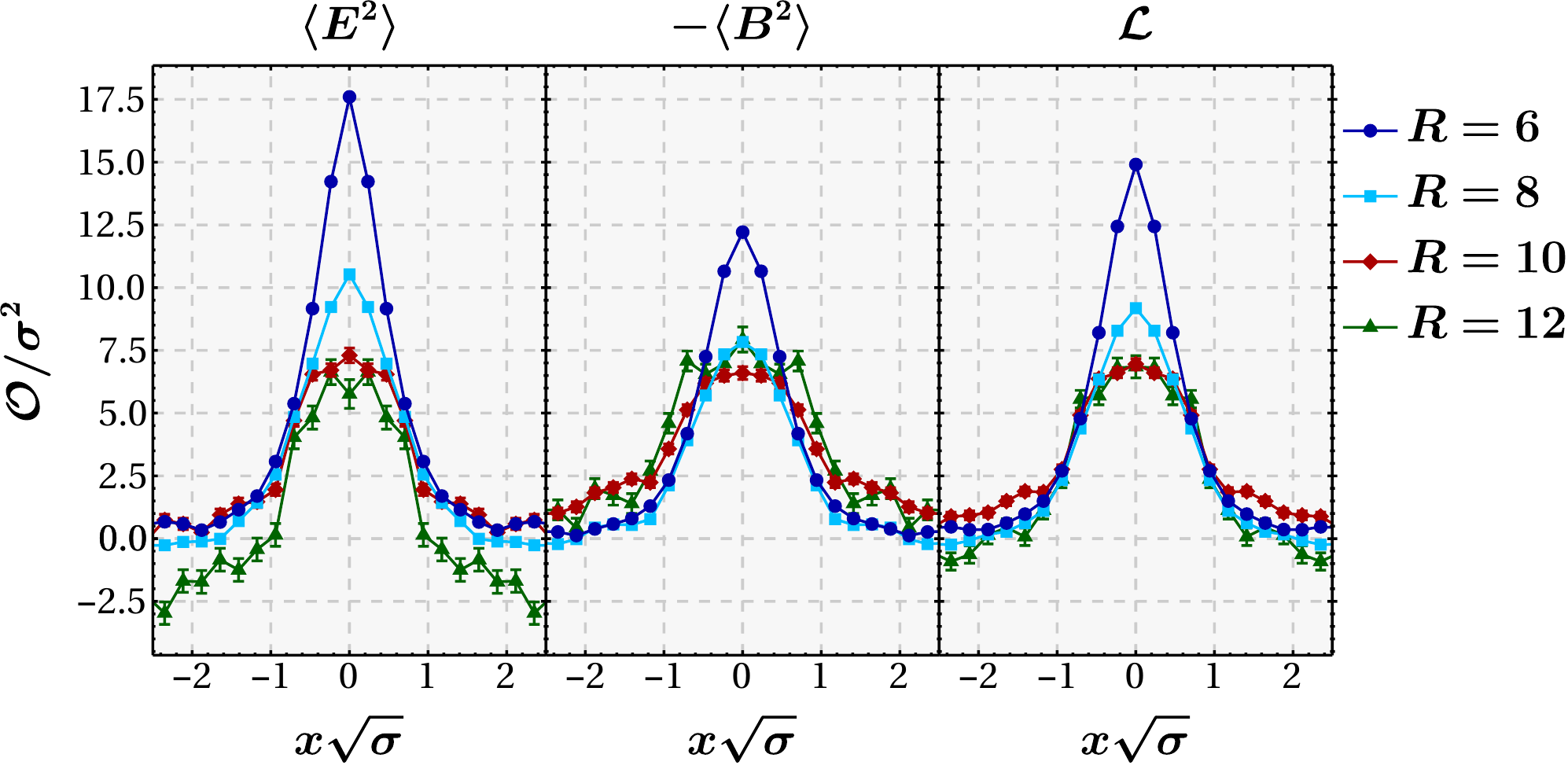}
\par\end{centering}}
\\
\subfloat[squared densities in the charge axis at $T=0.986\,T_c$.\label{fig:Field_XY_6p0534}]{
\begin{centering}
\includegraphics[width=8cm]{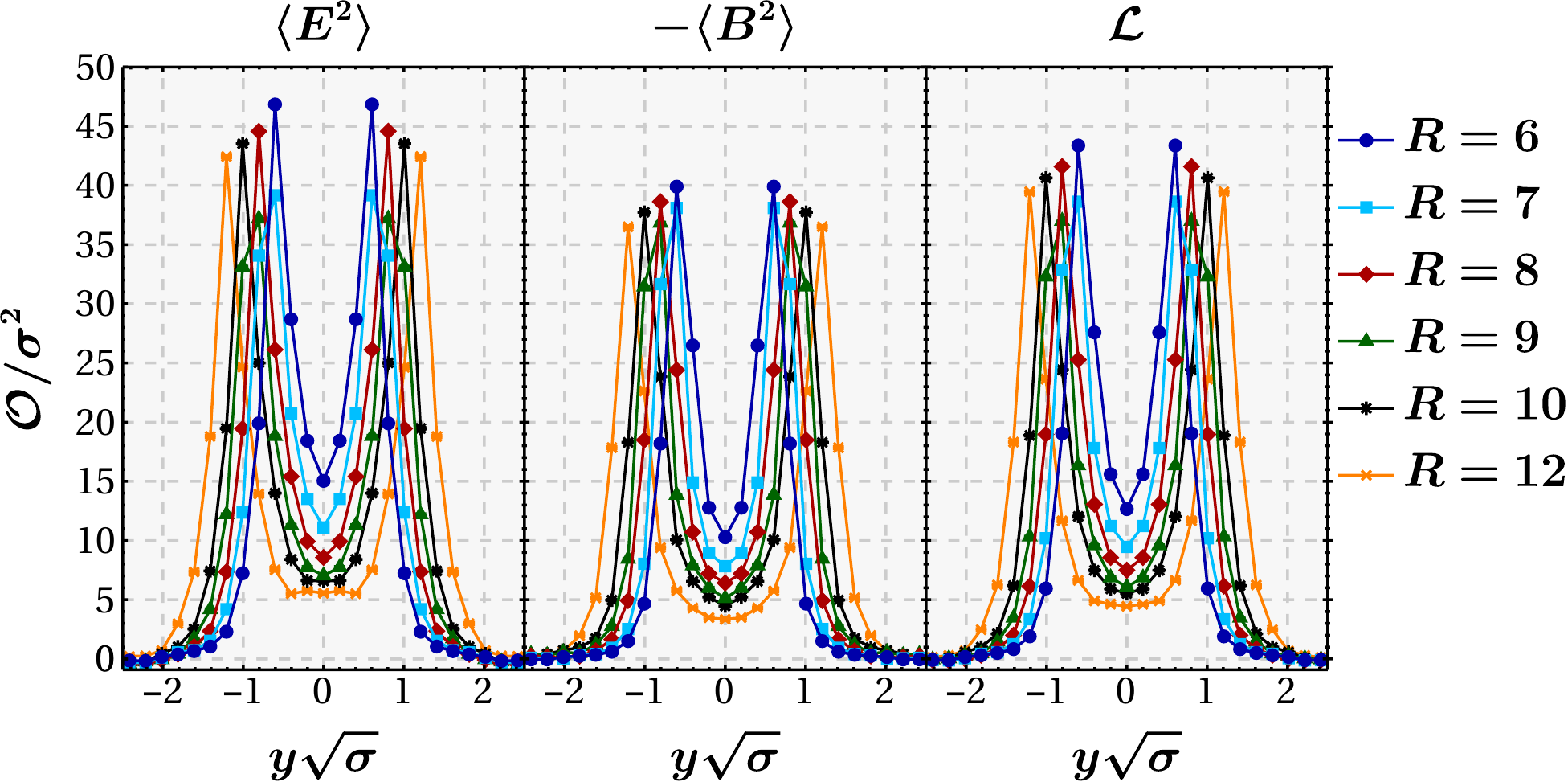}
\par\end{centering}}
\subfloat[squared densities in the mediator plane at $T=0.986\,T_c$.\label{fig:Field_XZ_6p0534}]{
\begin{centering}
\includegraphics[width=8cm]{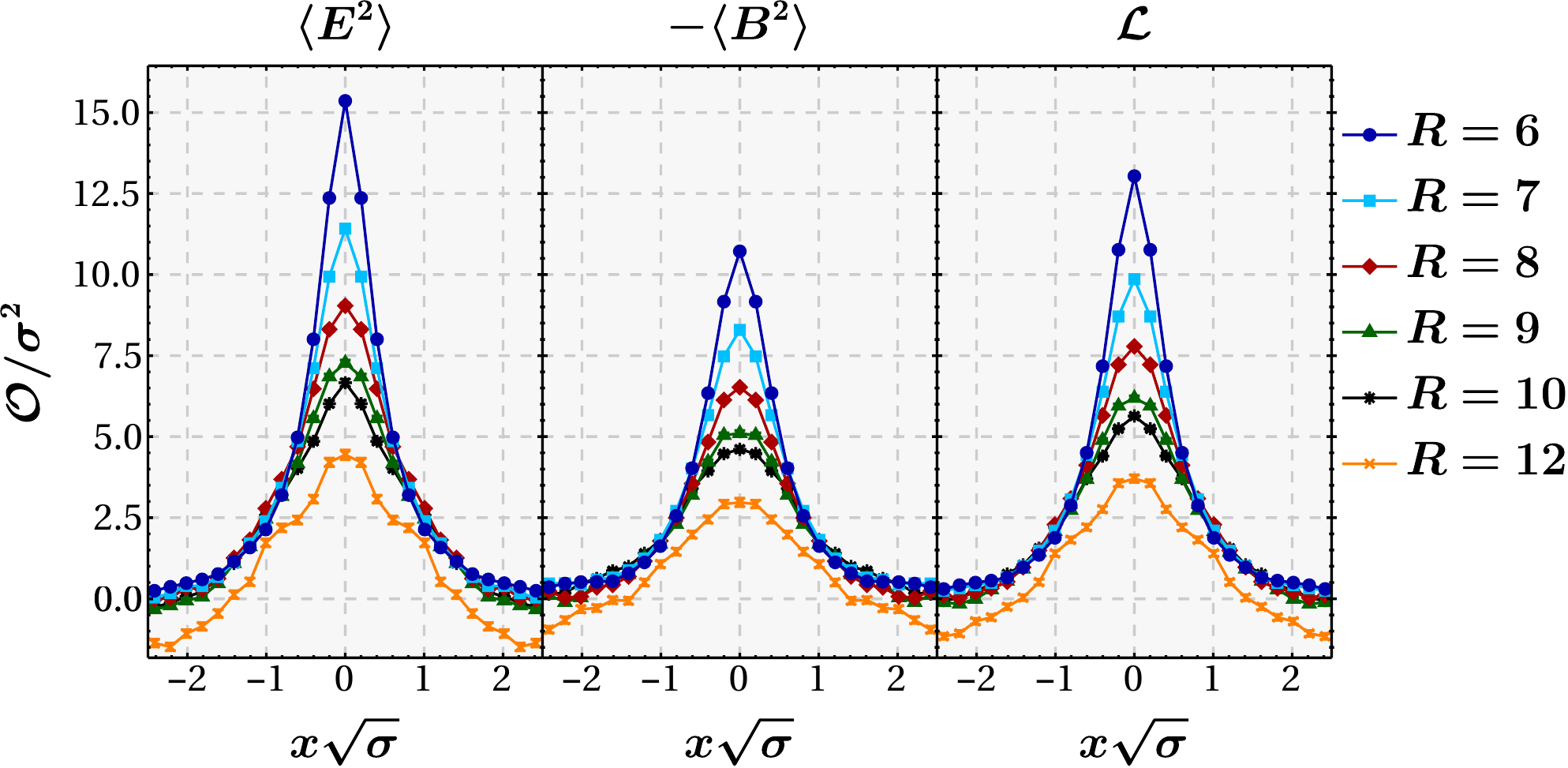}
\par\end{centering}}
\par\end{centering}
\caption{ (Colour Online.) Results for the chromoelectric field, chromomagnetic field and action density for the $Q\bar{Q}$ system at $T \leq T_c$.}
\label{fig:profiles_t0_ppdagger_fit}
\end{figure*}

\begin{figure*}[!t]
\begin{centering}
\captionsetup[subfloat]{farskip=0.1pt,captionskip=0.1pt}
\subfloat[squared densities in the charge axis at $T=1.127\,T_c$.\label{fig:Field_XY_6p13931}]{
\begin{centering}
\includegraphics[width=8cm]{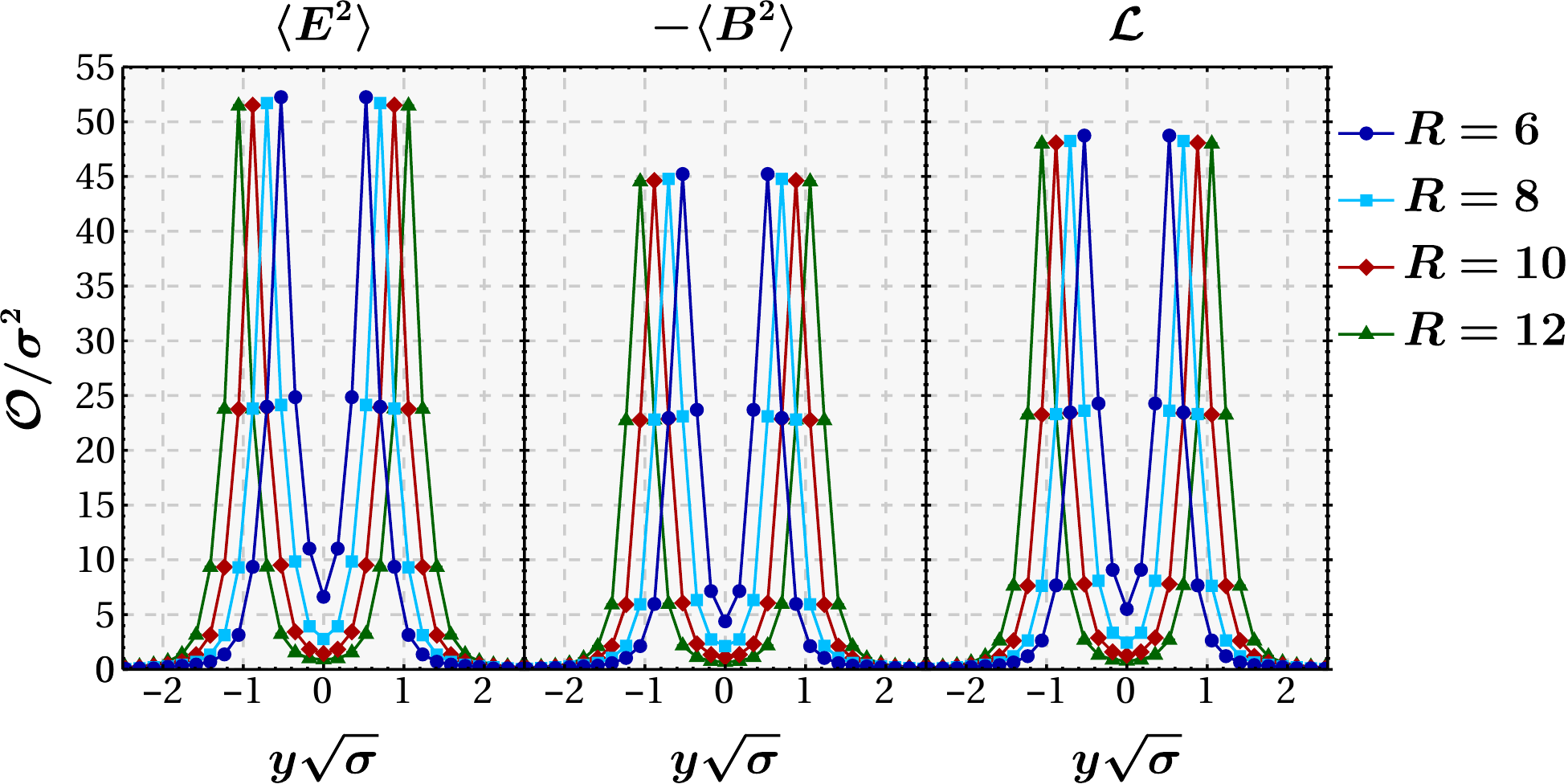}
\par\end{centering}}
\subfloat[squared densities in the mediator plane at $T=1.127\,T_c$.\label{fig:Field_XZ_6p13931}]{
\begin{centering}
\includegraphics[width=8cm]{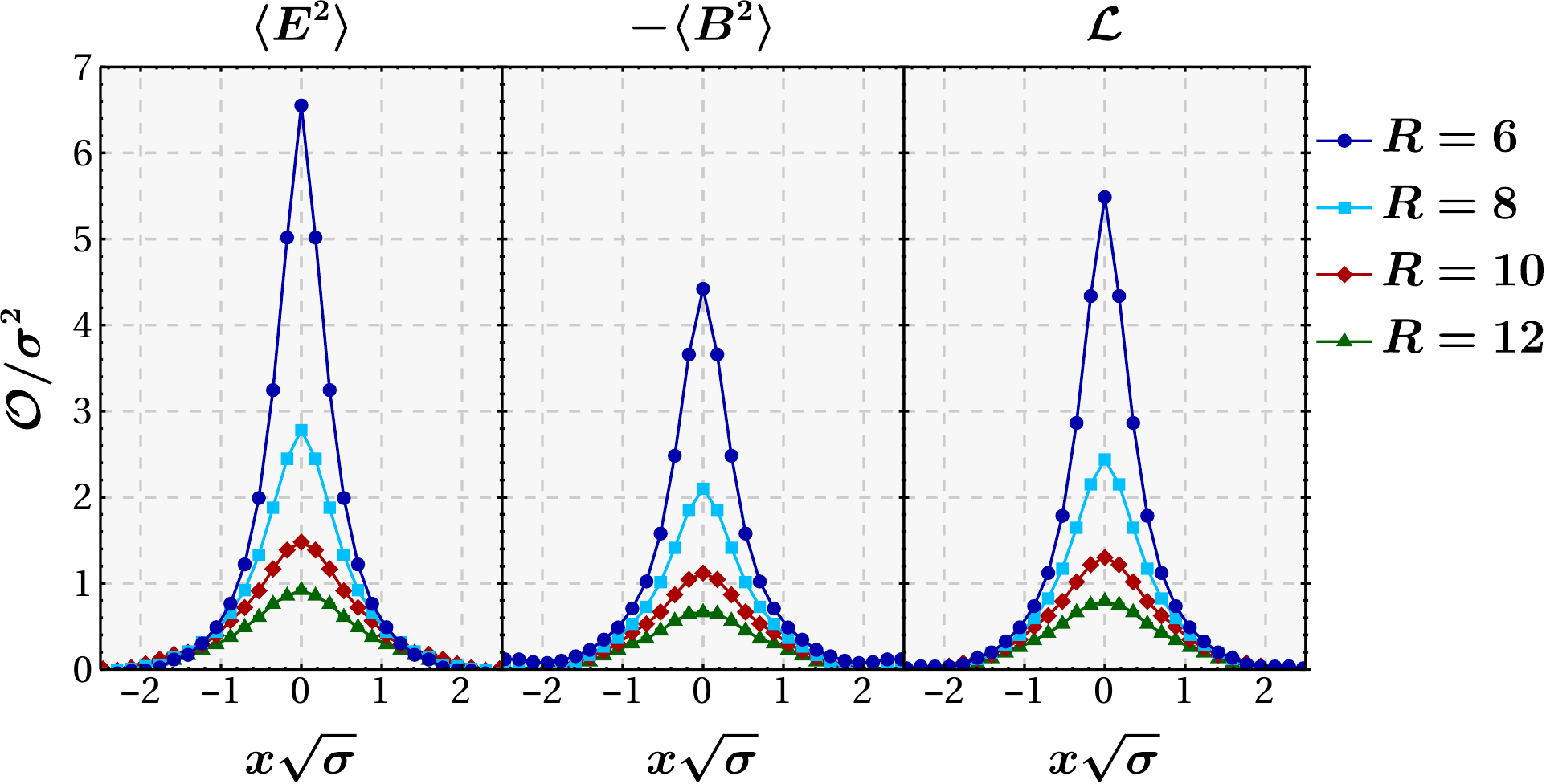}
\par\end{centering}}
\\
\subfloat[squared densities in the charge axis at $T=1.408\,T_c$.\label{fig:Field_XY_6p29225}]{
\begin{centering}
\includegraphics[width=8cm]{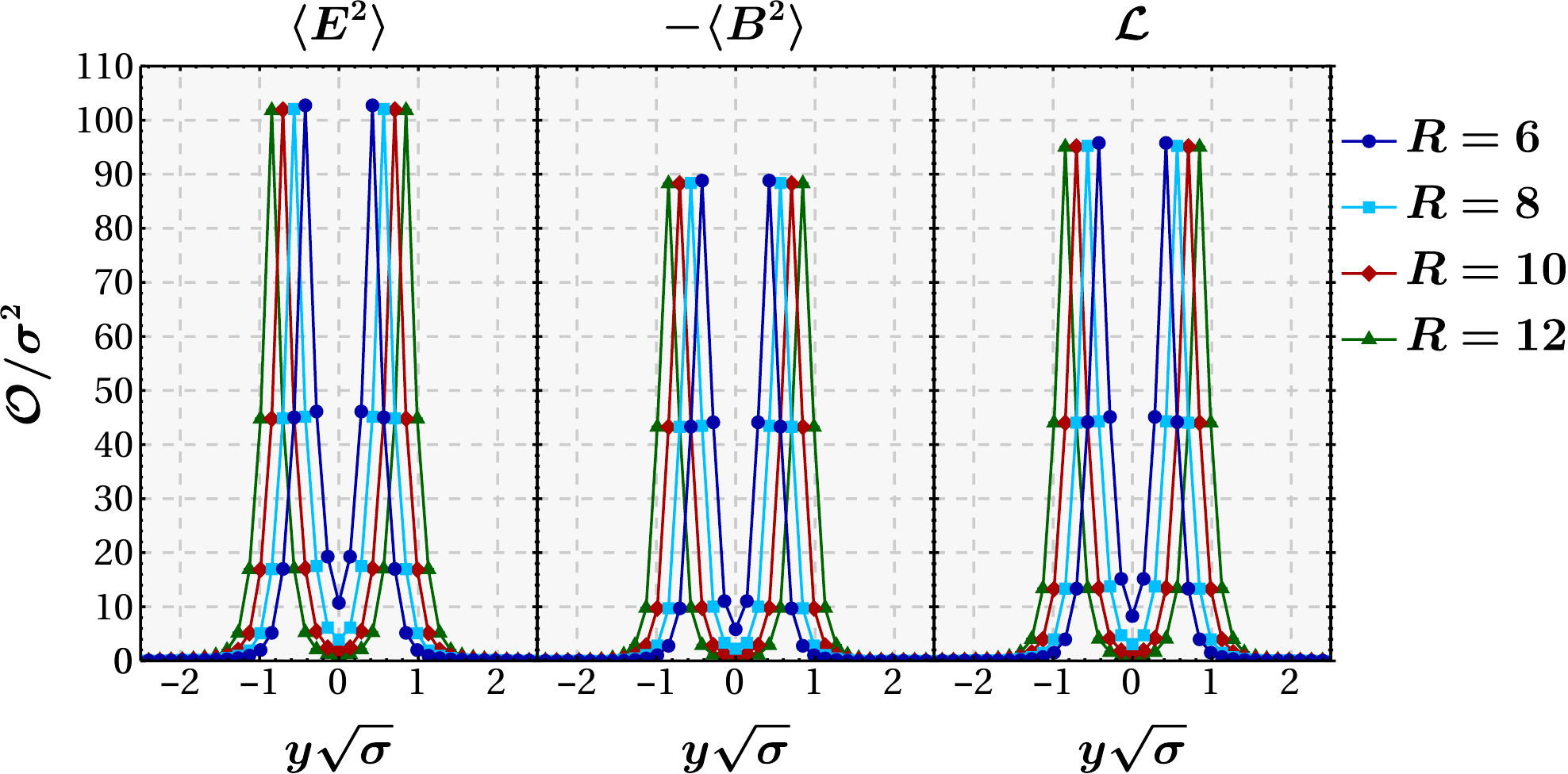}
\par\end{centering}}
\subfloat[squared densities in the mediator plane at $T=1.408\,T_c$.\label{fig:Field_XZ_6p29225}]{
\begin{centering}
\includegraphics[width=8cm]{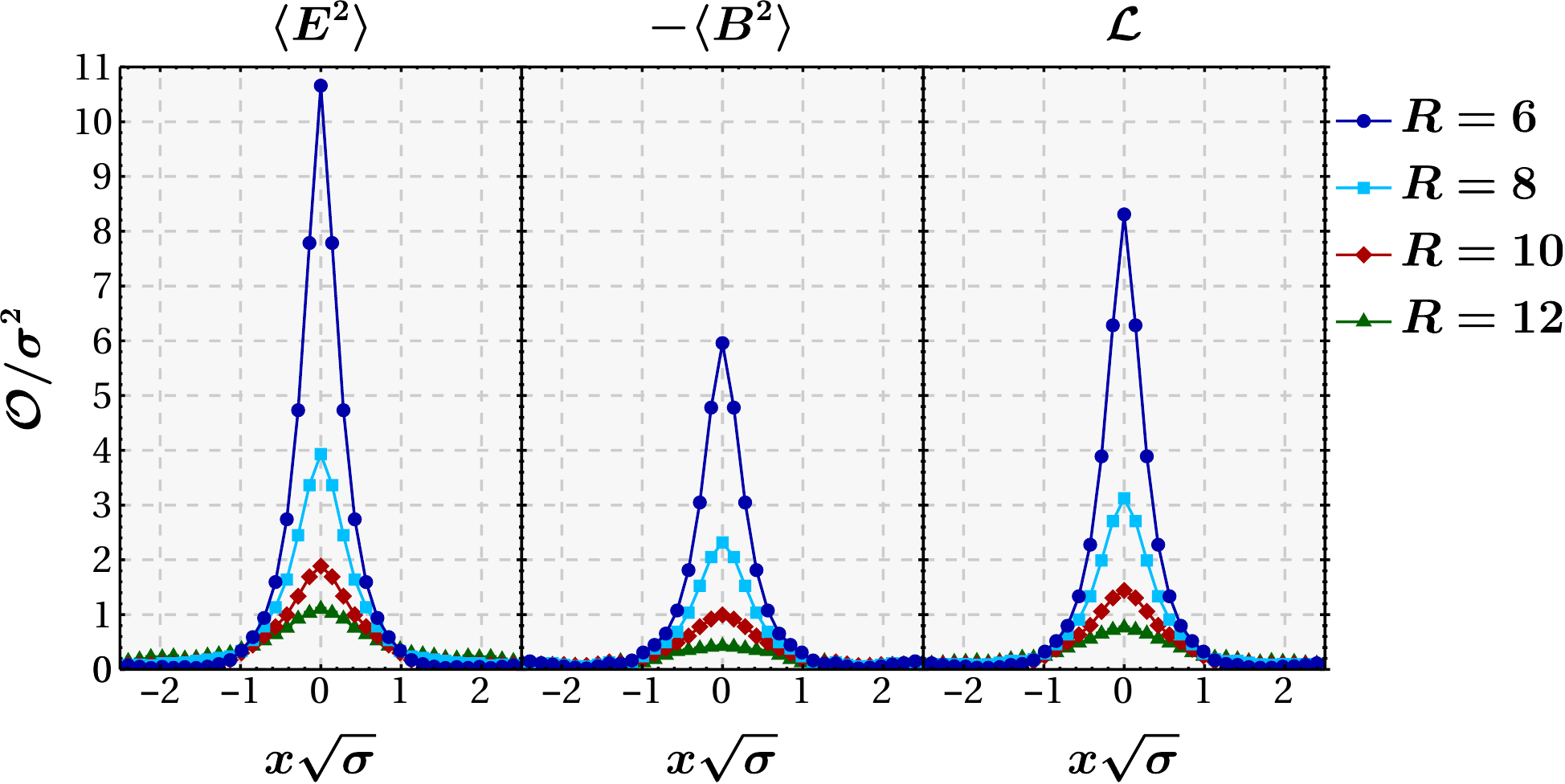}
\par\end{centering}}
\\
\subfloat[squared densities in the charge axis at $T=1.690\,T_c$.\label{fig:Field_XY_6p4249}]{
\begin{centering}
\includegraphics[width=8cm]{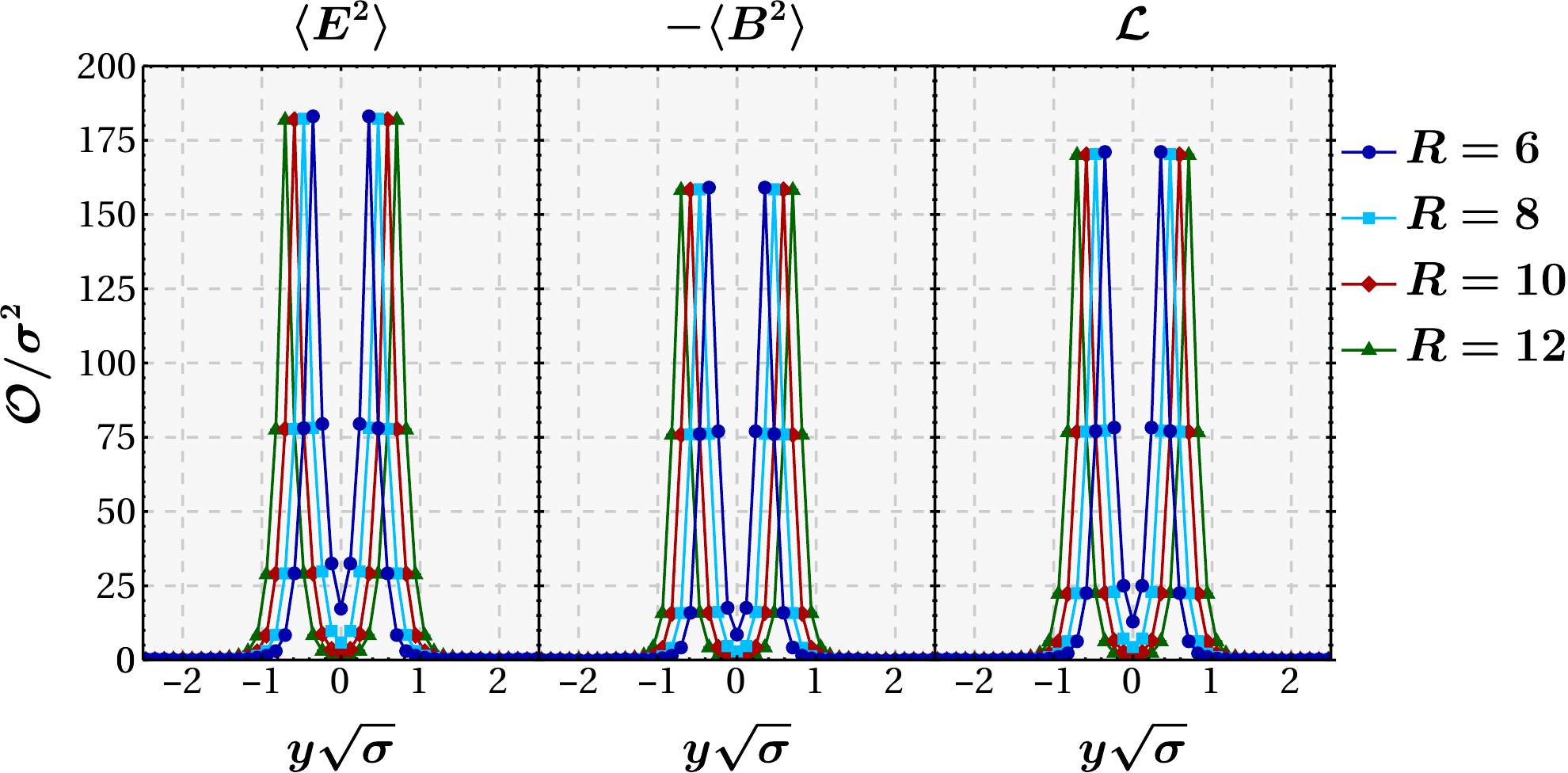}
\par\end{centering}}
\subfloat[squared densities in the mediator plane at $T=1.690\,T_c$.\label{fig:Field_XZ_6p4249}]{
\begin{centering}
\includegraphics[width=8cm]{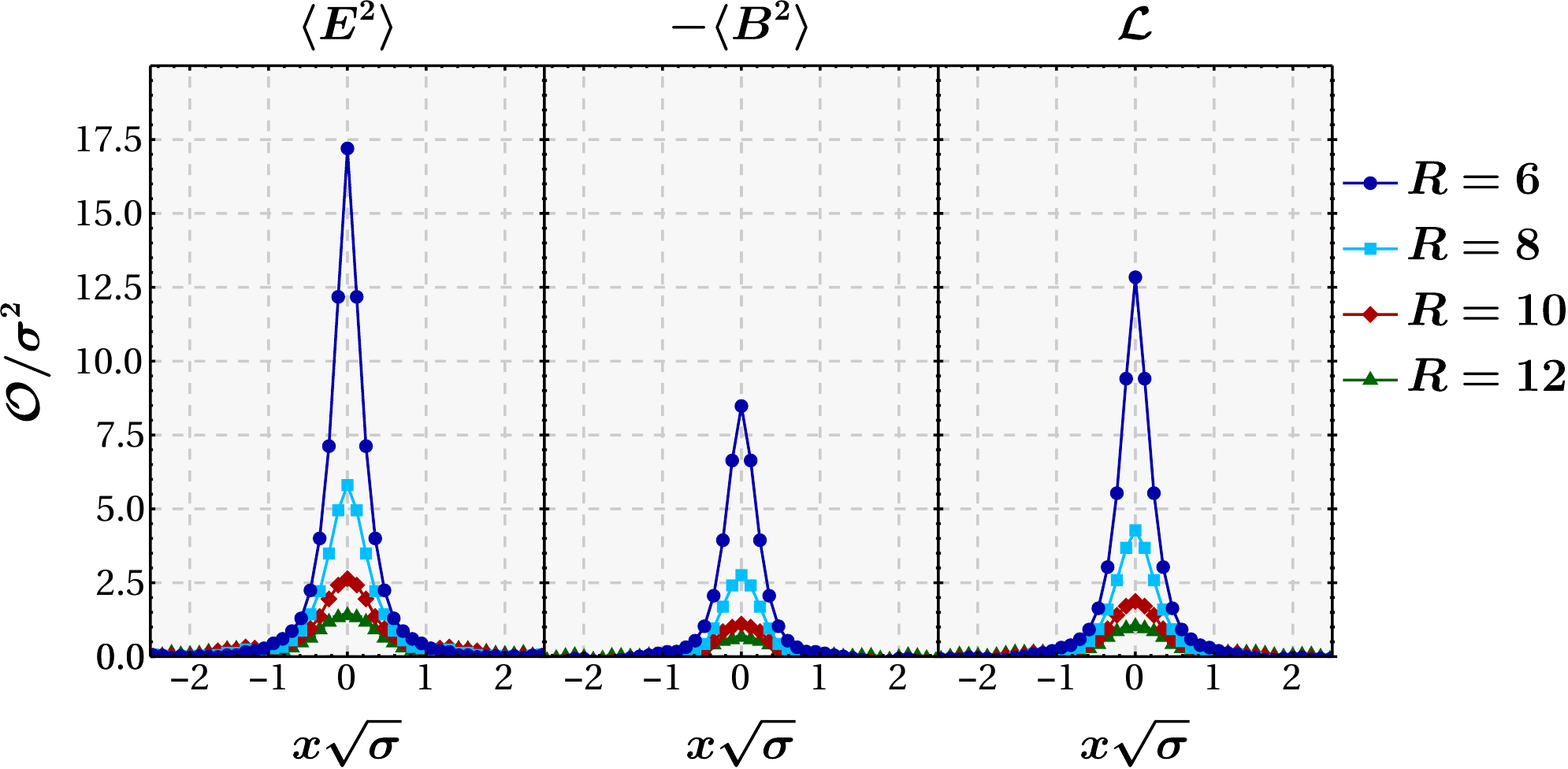}
\par\end{centering}}
\par\end{centering}
\caption{ (Colour Online.) Results for the chromoelectric field, chromomagnetic field and action density for the $Q\bar{Q}$ system at $T > T_c$.}
\label{fig:profiles_t1_ppdagger_fit}
\end{figure*}

\begin{figure*}[!t]
\captionsetup[subfloat]{farskip=0.1pt,captionskip=0.1pt}
\begin{centering}
\subfloat[squared densities in the charge axis at $T=1.127\,T_c$.\label{fig:Field_XY_6p13931_pp}]{
\begin{centering}
\includegraphics[width=8cm]{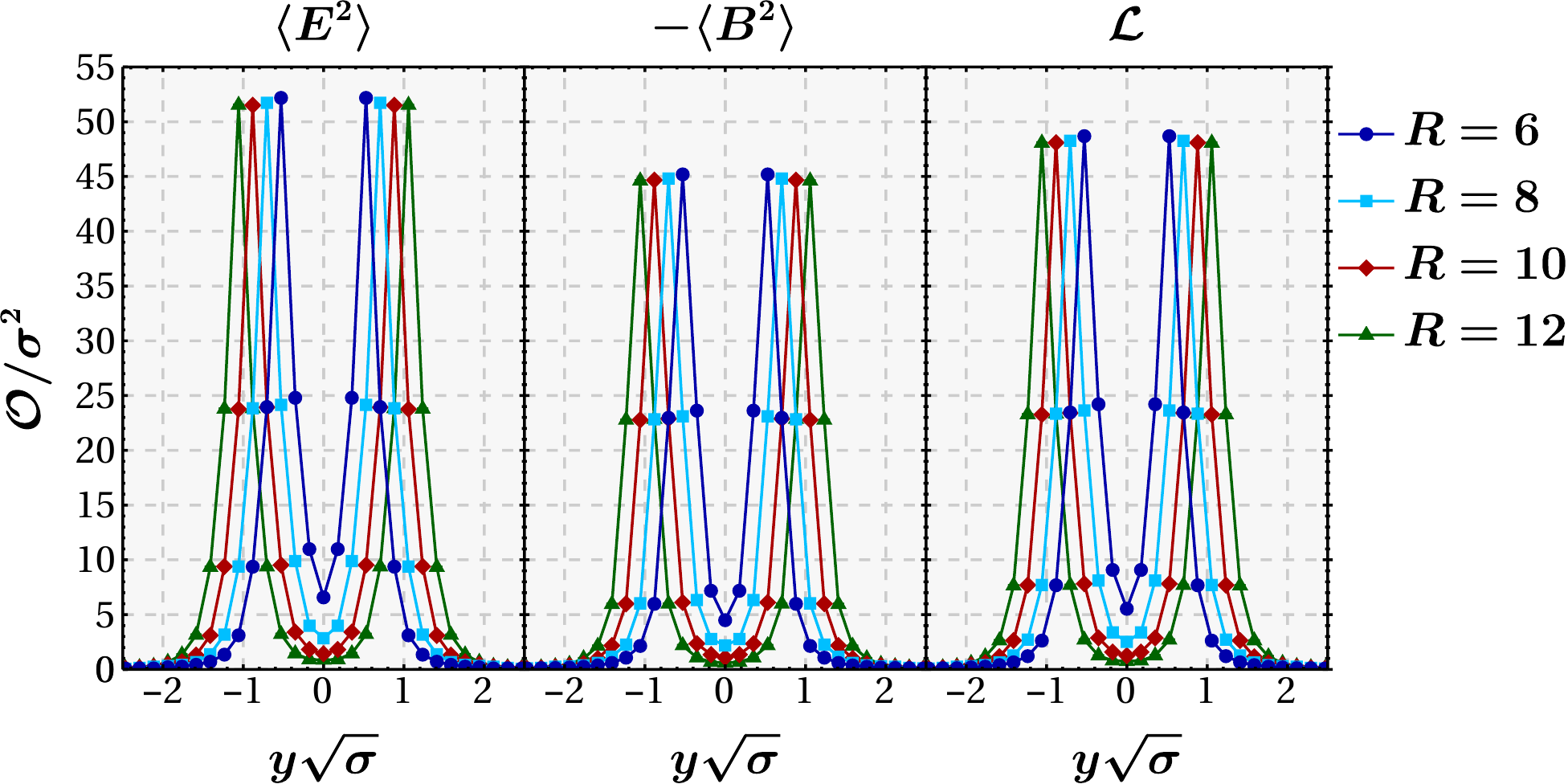}
\par\end{centering}}
\subfloat[squared densities in the mediator plane at $T=1.127\,T_c$.\label{fig:Field_XZ_6p13931_pp}]{
\begin{centering}
\includegraphics[width=8cm]{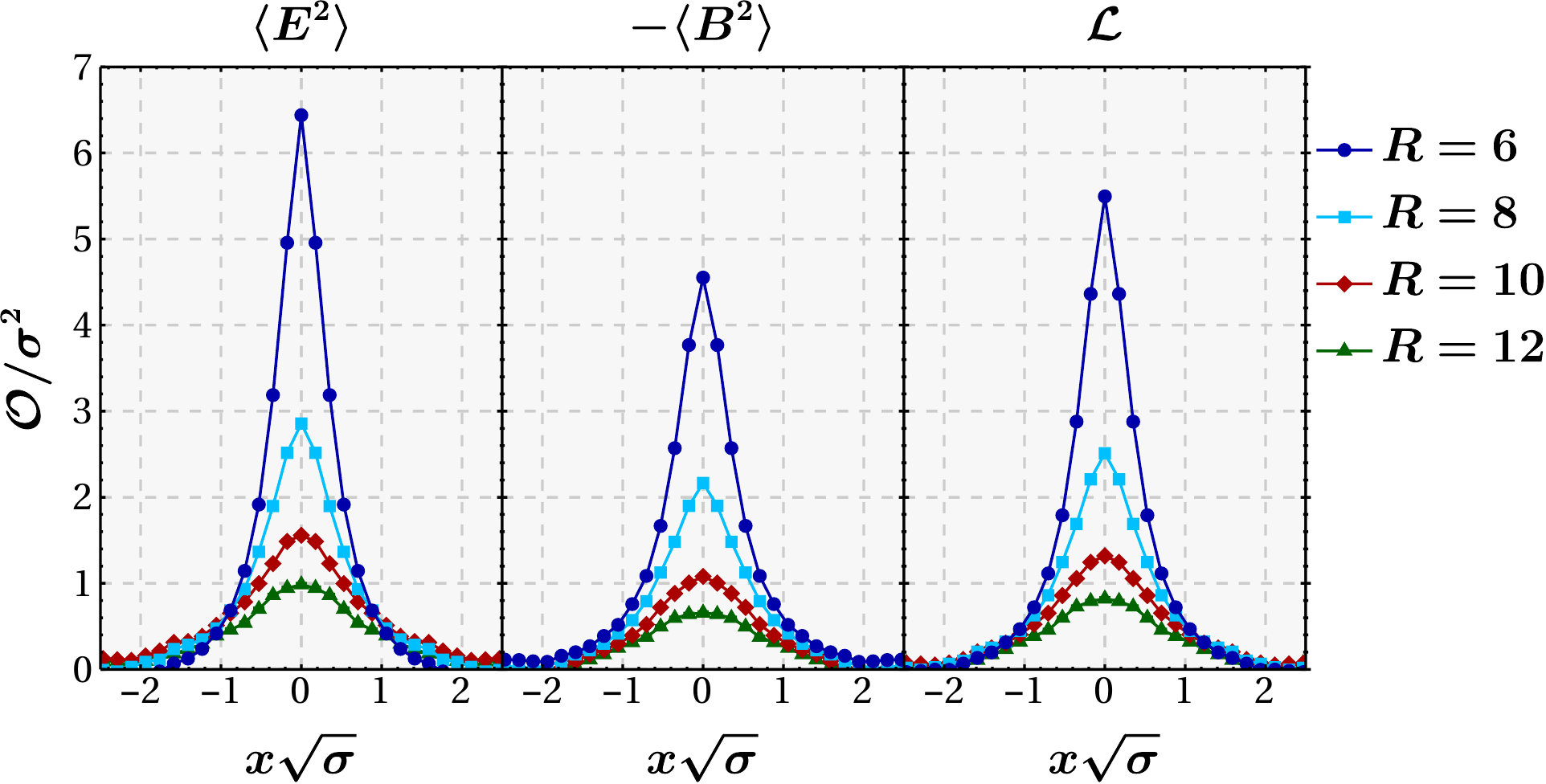}
\par\end{centering}}
\\
\subfloat[squared densities in the charge axis at $T=1.408\,T_c$.\label{fig:Field_XY_6p29225_pp}]{
\begin{centering}
\includegraphics[width=8cm]{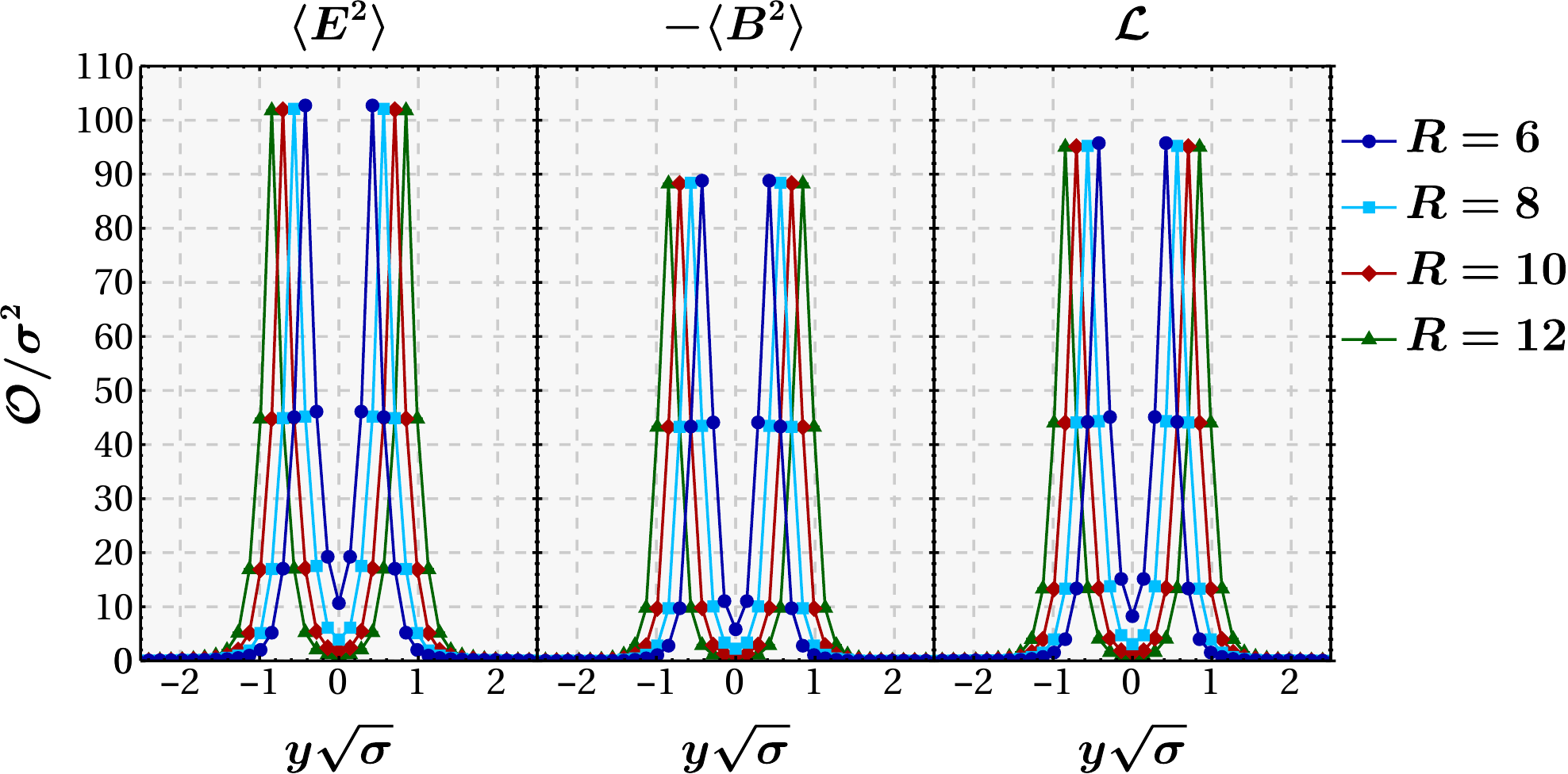}
\par\end{centering}}
\subfloat[squared densities in the mediator plane at $T=1.408\,T_c$.\label{fig:Field_XZ_6p29225_pp}]{
\begin{centering}
\includegraphics[width=8cm]{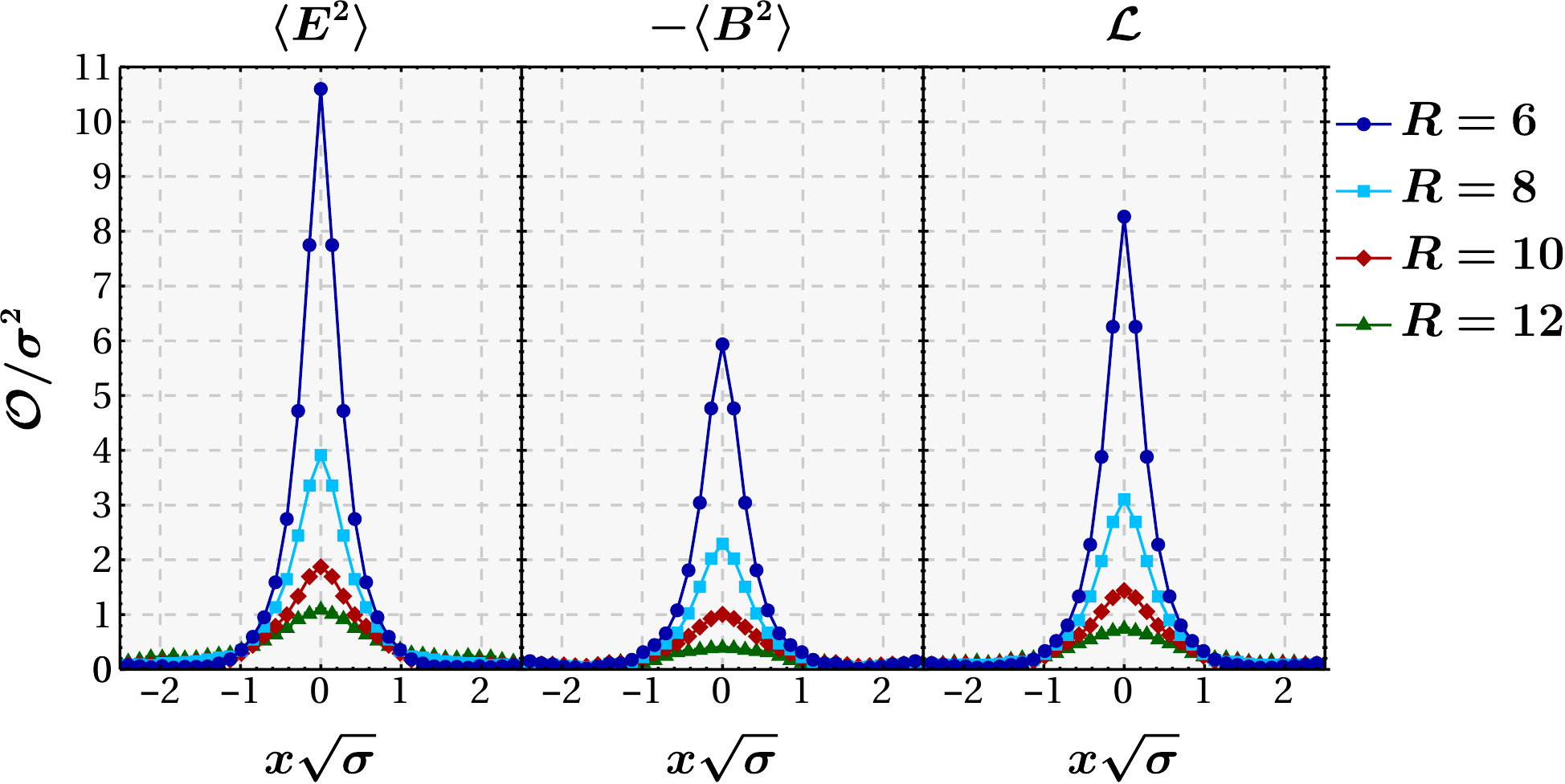}
\par\end{centering}}
\\
\subfloat[squared densities in the charge axis at $T=1.690\,T_c$.\label{fig:Field_XY_6p4249_pp}]{
\begin{centering}
\includegraphics[width=8cm]{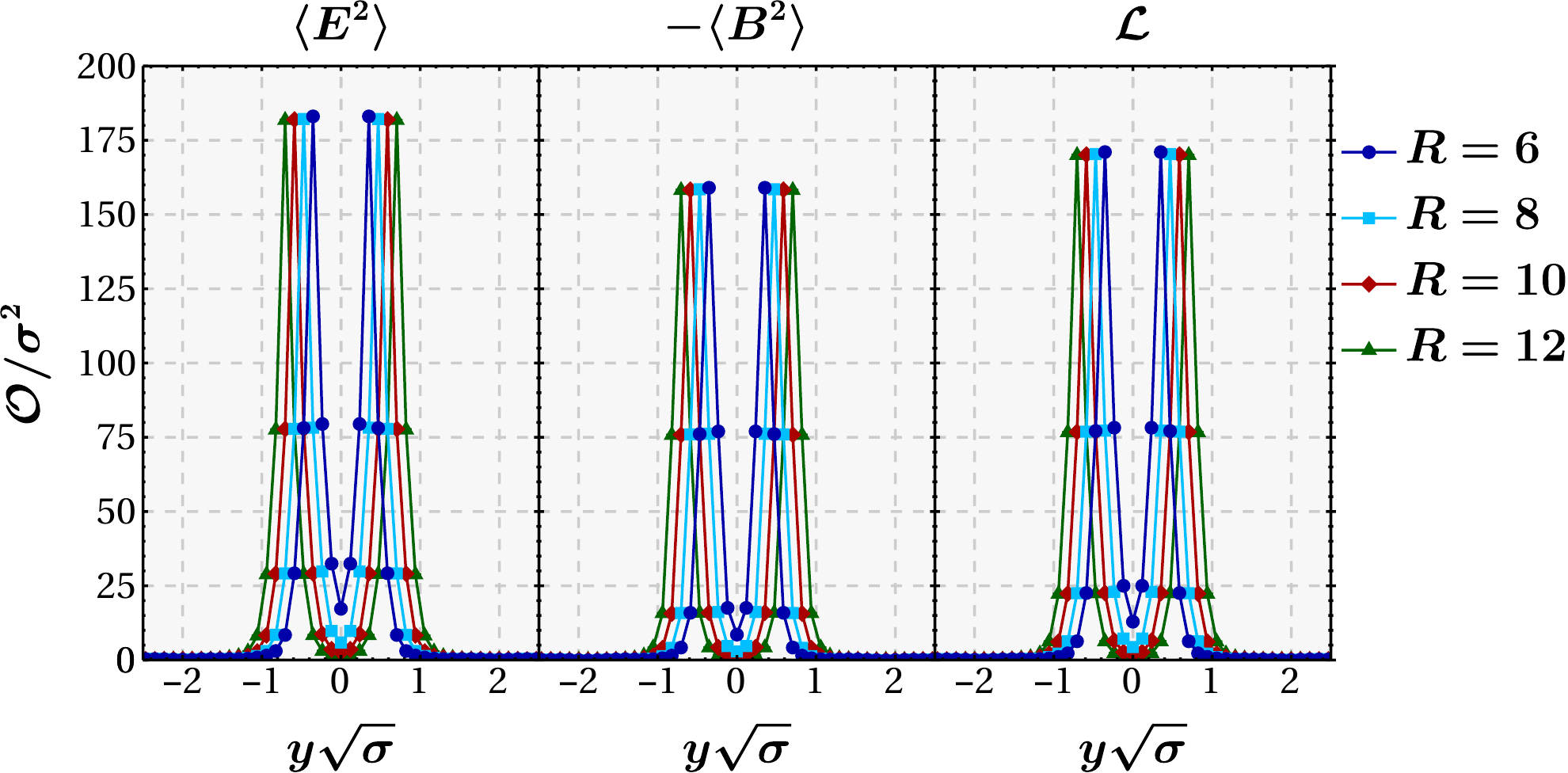}
\par\end{centering}}
\subfloat[squared densities in the mediator plane at $T=1.690\,T_c$.\label{fig:Field_XZ_6p4249_pp}]{
\begin{centering}
\includegraphics[width=8cm]{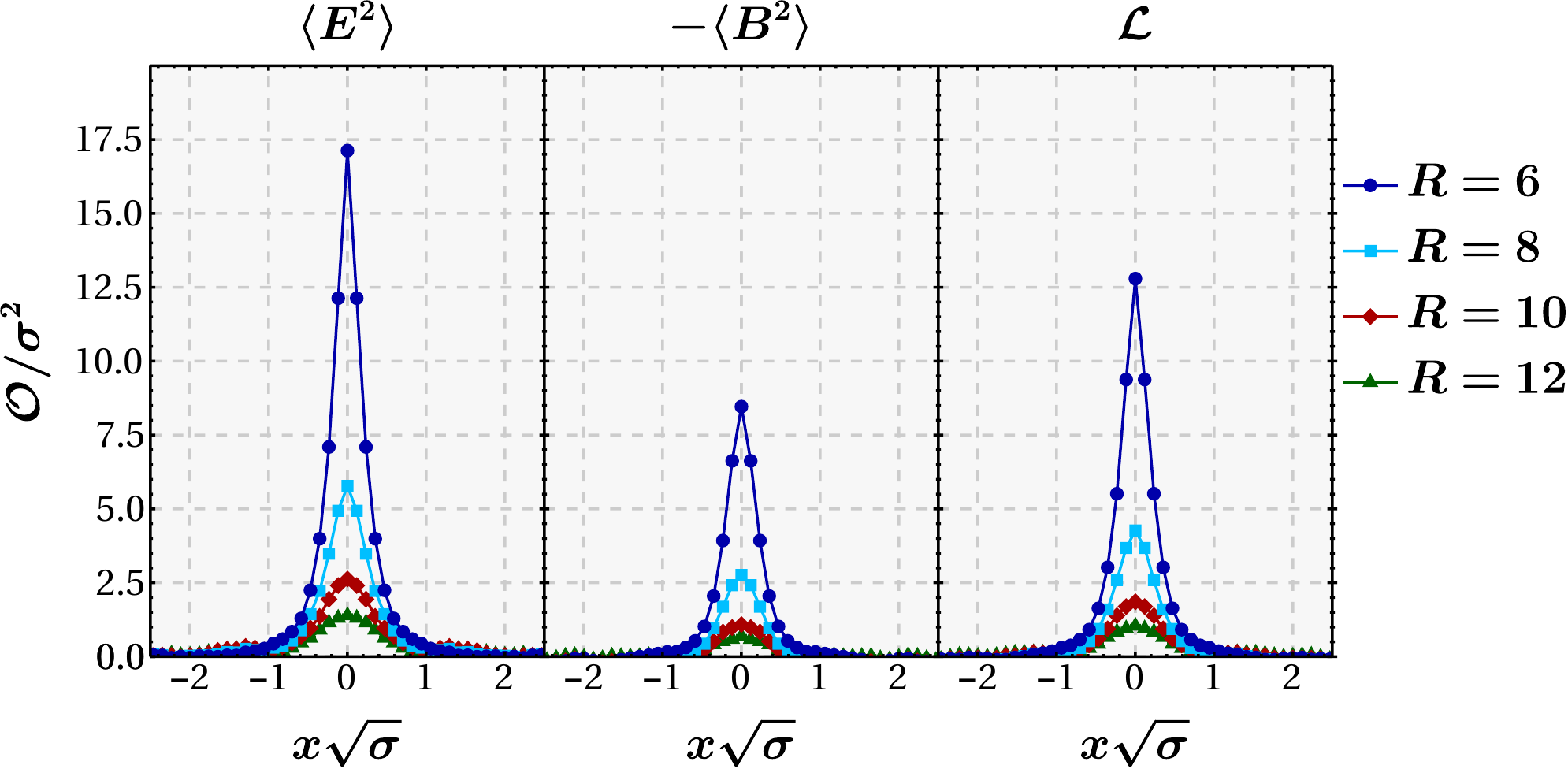}
\par\end{centering}}
\par\end{centering}
\caption{ (Colour Online.) Results for the chromoelectric field, chromomagnetic field and action density for the $QQ$ system.}
\label{fig:profiles_pp}
\end{figure*}

\begin{figure}[!t]
\includegraphics[width=0.95\columnwidth]{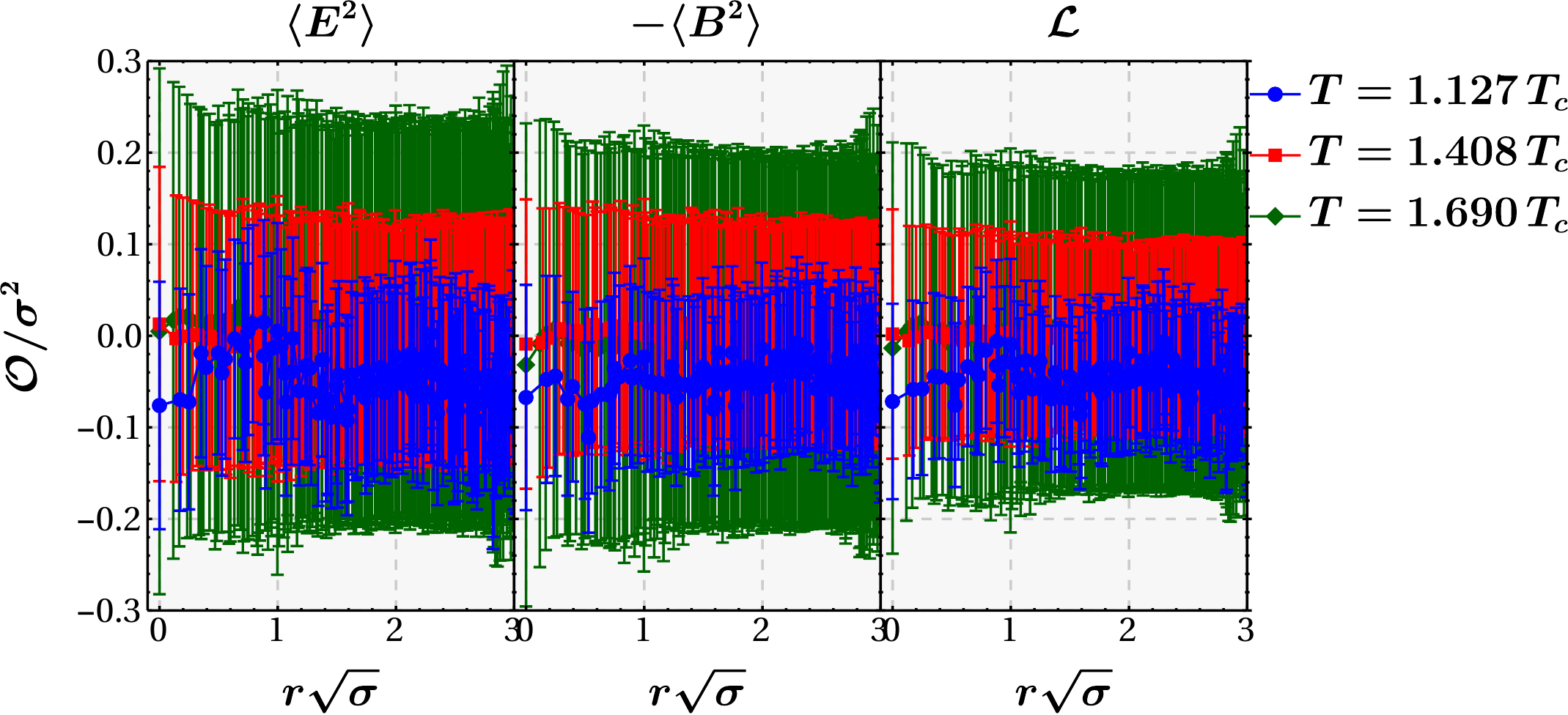}
\caption{
\color{black}
 (Colour Online.) Difference of the profiles for the $Q Q$ and $Q \bar Q$ systems, at our three temperatures $T > T_c$. The difference is consistent with zero modulo the statistical error bars, i e the two field densities are essentially identical.}
\label{fig:difference}
\end{figure}

\begin{figure*}[!t]
\hspace{-15pt}
\begin{centering}
\includegraphics[width=0.66\columnwidth]{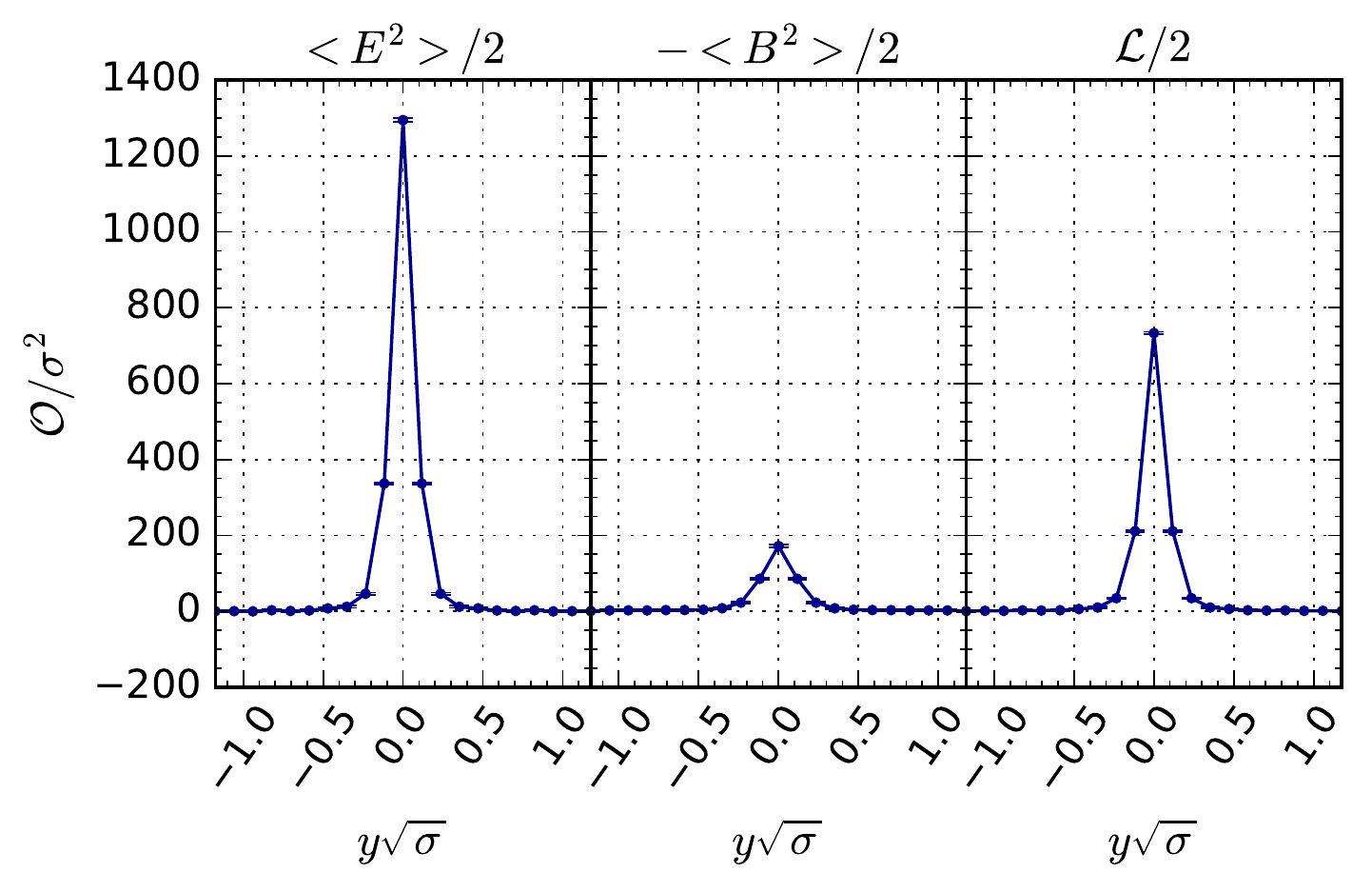}
\includegraphics[width=0.66\columnwidth]{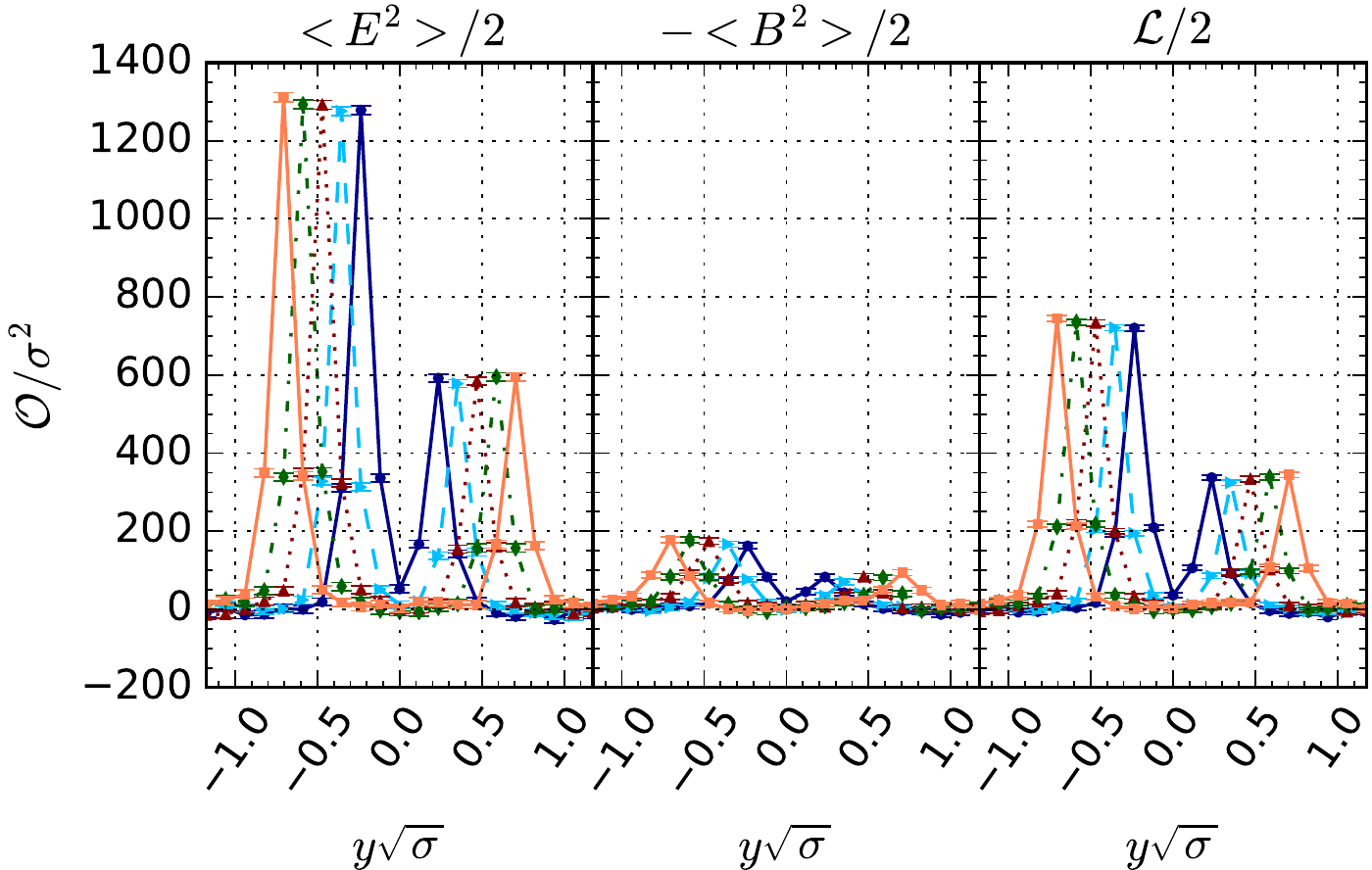}
\includegraphics[width=0.66\columnwidth]{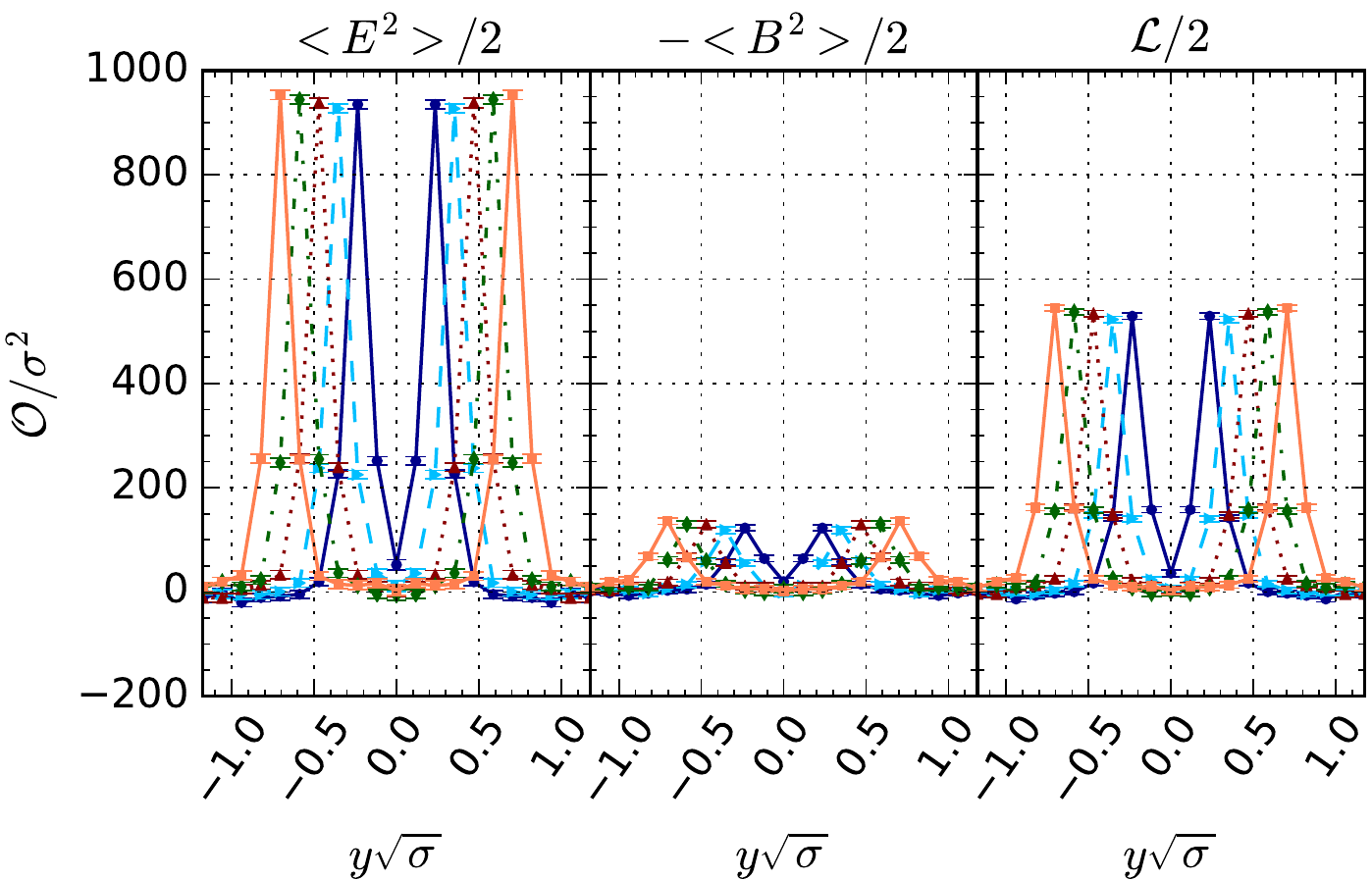}
\par\end{centering}
\vspace{5pt}
\par
\hfil \hfil \hfil \includegraphics[width=0.80\columnwidth]{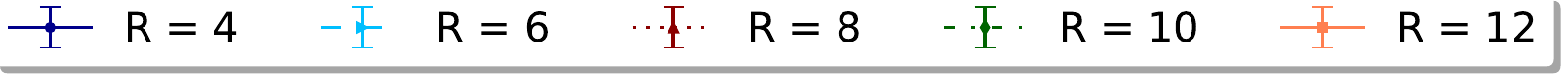}
\caption{ (Colour Online.) Charge axis squared field densities for the (left) single adjoint source $A$ system, (centre) adjoint source-quark $QA$ system and (right) adjoint source-adjoint source $AA$ system, all for $\beta=6.4249$, $T=1.690T_c$.}
\label{fig:GandQGsystems}
\end{figure*}

Therefore, using the plaquette orientations $(\mu,\nu)=(2,3), (3,1), (1,2),$ $(1,4),(2,4),(3,4)$, we can relate the six components in \cref{eq:fmunucomp} to the components of the chromoelectric and chromomagnetic fields,
\begin{equation}
f_{\mu\nu}\rightarrow\frac{1}{2}\left(-\Braket{B_x^2},-\Braket{B_y^2},-\Braket{B_z^2},\Braket{E_x^2},\Braket{E_y^2},\Braket{E_z^2}\right) \ ,
\end{equation}
and also calculate the total action (Lagrangian) density, 
\begin{equation}
\Braket{\mathcal{L}}=\frac12\left(\Braket{E^2}-\Braket{B^2}\right) \ .
\label{eq:lagrangiand}
\end{equation}

In order to improve the signal over noise ratio in the $Q\bar{Q}$ and $QQ$ systems, we use the extended multihit technique detailed in Ref. \cite{Cardoso:2013lla}, which is an extended version of the multihit technique  \cite{Brower:1981vt, Parisi:1983hm}.
The technique consists in replacing each temporal link by its thermal average with the first $N$th order neighbours fixed, whereas the simple multihit would just take the thermal average of a temporal link with the first neighbours. We apply the heat-bath algorithm to all the links inside, averaging the central link,
\begin{equation}
U_4\rightarrow \bar{U}_4=\frac{\int \left[\mathcal{D}U_4\right]_\Omega U_4 \,e^{\beta \sum_{\mu,s} \Tr\left[ U_\mu(s) F^\dagger(s)\right]}}{\int \left[\mathcal{D}U_4\right]_\Omega \,e^{\beta \sum_{\mu,s} \Tr\left[ U_\mu(s) F^\dagger(s)\right]}} \ .
\end{equation}
By using $N = 2$ we are able to greatly improve the signal, when compared with the error reduction achieved with the simple multihit. Of course, this technique is more computer intensive than simple multihit, while being simpler to implement than multilevel
\cite{Luscher:2001up}.
The only restriction is $R > 2N$ for this technique to be valid.

Moreover, just below the phase transition, we need to make sure that we don't have contaminated configurations as already mentioned in \cite{Cardoso:2011hh}. By plotting the histogram of  Polyakov loop history for $\beta=6.055$ we are able to identify a second peak, see \cref{fig:histpolyloop}. Thus we remove all the configurations that lie on the second peak \cite{Cardoso:2011hh}. Therefore, in \cref{tab:latticesimdata} the value with asterisk corresponds to the configurations after removing these contaminated configurations. 

\section{Results for the squared field densities in the charge axis and in the mediator plane}

In this section, we present the results for different $\beta$ values using a fixed lattice size of $48^3\times 8$. Our diferent ensembles are detailed in \cref{tab:latticesimdata}. We compute the lattice spacing, 
 in units of the string tension at zero temperature, using the parametrization of Ref.  \cite{Edwards:1997xf}. 
 All our computations are fully performed in NVIDIA GPUs using our CUDA language.

The two charges, $Q$ $\bar{Q}$ or $A$ (adjoint charge), are located at coordinates $(0,0,-R/2,0)$ and $(0,0,R/2,0)$, illustrated in Fig. \ref{fig:geometryFiniteT},  for $R$ between 4 and 14 in lattice spacing units at $T$=0, and at finite $T$ we start at $R=6$ and stop at $R=12$. For $\beta=6.0534$, we also study odd values, $R=7 \text{ and } 9$. 

{
\color{black}

Our results consist in the squared field densities in two spacial subspaces, the charge axis ($z$ axis) including the two charges, and the mediator plane (plane $x,y$) of the two charges.

\subsection{$Q \bar Q$ flux tube at $T < T_c$}

In \cref{fig:profiles_t0_ppdagger_fit}, we show the flux tube squared field densities $\langle E^2 \rangle$, $-\langle B^2 \rangle$, $ \langle \cal L \rangle$ for the $Q\bar{Q}$ system at temperatures $T < T_c$.  In the left sub-figures we show the charge axis and in the right sub-figures we show the mediator plane. The top sub-figures correspond to the temperature $T=0.845 T_c$, and the bottom sub-figures correspond to $T=0.0.986 T_c$.

As a first qualitative analysis, since confinement should become weaker for higher temperatures, the flux tubes should become less squeezed. We expect the flux tubes to be less dense at higher temperatures. Indeed in the right panels of \cref{fig:profiles_t0_ppdagger_fit}, it is clear the intensity of the fields does decrease with the temperature.

\subsection{$Q \bar Q$ field densities at $T > T_c$}

In \cref{fig:profiles_t1_ppdagger_fit}, we show the results for the $Q\bar{Q}$ system at temperatures at $T > T_c$. As in \cref{fig:profiles_t0_ppdagger_fit}, we show the flux tube squared field densities $\langle E^2 \rangle$, $-\langle B^2 \rangle$, $ \langle \cal L \rangle$.  In the left sub-figures we show the charge axis and in the right sub-figures we show the mediator plane. Now the top sub-figures correspond to temperature $T=1.127\,T_c$, the middle sub-figures correspond to temperature $T=1.408\,T_c$ and the bottom sub-figures correspond to $T=1.690 T_c$.

What is clear now is that the intensity of the flux tube does decrease while the inter-charge distance increases. This is visible in the right pannels of \cref{fig:profiles_t1_ppdagger_fit}, in a behaviour different from the equivalent sub-figures of \cref{fig:profiles_t0_ppdagger_fit}. This suggests the flux tubes no longer exist above the deconfinement temperature $T_c$.


\subsection{$Q \bar Q$ field densities at $T > T_c$}

In \cref{fig:profiles_pp}, we show the results for the $Q Q$ system at $T > T_c$ (below $T_c$ the Polyakov loop of non colour-singlet systems vanish). For an easier comparison with the $Q \bar Q$ system, the six different sub-figures are ordered exactly as in \cref{fig:profiles_t1_ppdagger_fit}.

 It is remarkable that the field densities of the $Q Q$ system in \cref{fig:profiles_pp} are apparently identical, modulo statistic errors, to the ones of the $Q\bar{Q}$ system in \cref{fig:profiles_t1_ppdagger_fit}. This similarity was not anticipated, and it may be relevant for the various QCD models based in Polyakov loops.

To check in more detail the difference, we plot in \cref{fig:difference} the difference between the squared field densities of the $Q Q$ and of the $Q \bar Q$ system at our three temperatures $T > T_c$. The difference is consistent with zero modulo the statistical error bars, i e the two field densities are essentially identical.
 
This may possibly be interpreted as an evidence for the uncorrelation of the different Polyakov loops at  temperatures $T > T_c$, and thus for the non-existence of a flux tube, which should be intrinsically non-linear. It is well known  \cite{Lo:2013hla}
 that, in the confined phase below $T_c$, the Polyakov loop has $Z3$ symmetry, in the sense it tends to take values close to the three cubic roots of 1, at $\left\{1, {-1 + \sqrt 3 \, i \over 2}, {-1 -\sqrt 3 \, i \over 2} \right\}$; and its average value vanishes. When we have a $Q \bar Q$ system, the fields correlate in a non-linear flux tube, and the average no longer vanishes for $ T< T_c$, whereas the Polyakov loops of a $ Q Q$ system vanish. Now, in the deconfined phase above Tc, the $Z3$ symmetry is broken, the Polyakov loop gets closer to 1, and the Polyakov loop average over all configurations is real. 
This is for instance utilized in matrix models for the deconfinement phase transition
\cite{Pisarski:2000eq,Pisarski:2006hz,Dumitru:2012fw,Smith:2013msa,Bicudo:2014cra}. 
 In this sense the Polyakov loop of a $Q$ is on average identical to its complex conjugate (see Eq. (\ref{eq:polyakov_loops}), the Polyakov loop of a $\bar Q$.
 Thus, in case there are no non-linear correlations between the two Polyakov loops present on the lattice, 
 it is plausible the field densities for the $Q Q $ and $Q \bar Q$ are identical.

}

\subsection{$Q A$ field densities at $T > T_c$}

Moreover, in \cref{fig:GandQGsystems} we study the effect of including a static adjoint source in the system with an adjoint Polyakov loop, as detailed in Eq. \ref{eq:polyakov_loops}. 

{\color{black}
In \cref{fig:GandQGsystems} we show the squared field densities $\langle E^2 \rangle$, $-\langle B^2 \rangle$, $ \langle \cal L \rangle$ for three systems, a single adjoint source $A$ system, the adjoint source-quark $QA$ system and the adjoint source-adjoint source $AA$ system, all for our highest temperature with $\beta=6.4249$, $T=1.690T_c$. This is just a first study, possibly interesting for the effective models of QCD with Polyakov loops, and we do not perform an analysis as detailed as in the $QQ$ and $Q \bar Q$ systems.
}

Nevertheless, as in the case with quark sources $QQ$ and $Q \bar Q$, the plots suggest the total square fields of the $QA$ system are approximately similar to a simple linear sum of the square fields produced by two charges. Again, we find this linear superposition  contradicts the existence of flux tubes at $T > T_c$, since flux tubes are clearly non-linear objects.

\section{Analysis of tube widening, including systematic errors}

\begin{figure}[!t]
\captionsetup[subfloat]{farskip=0.1pt,captionskip=0.1pt}
\begin{centering}
\subfloat[$Q\bar{Q}$.\label{fig:FieldSR1Da_1p410_XZ_ppdagger}]{
\begin{centering}
\includegraphics[width=8cm]{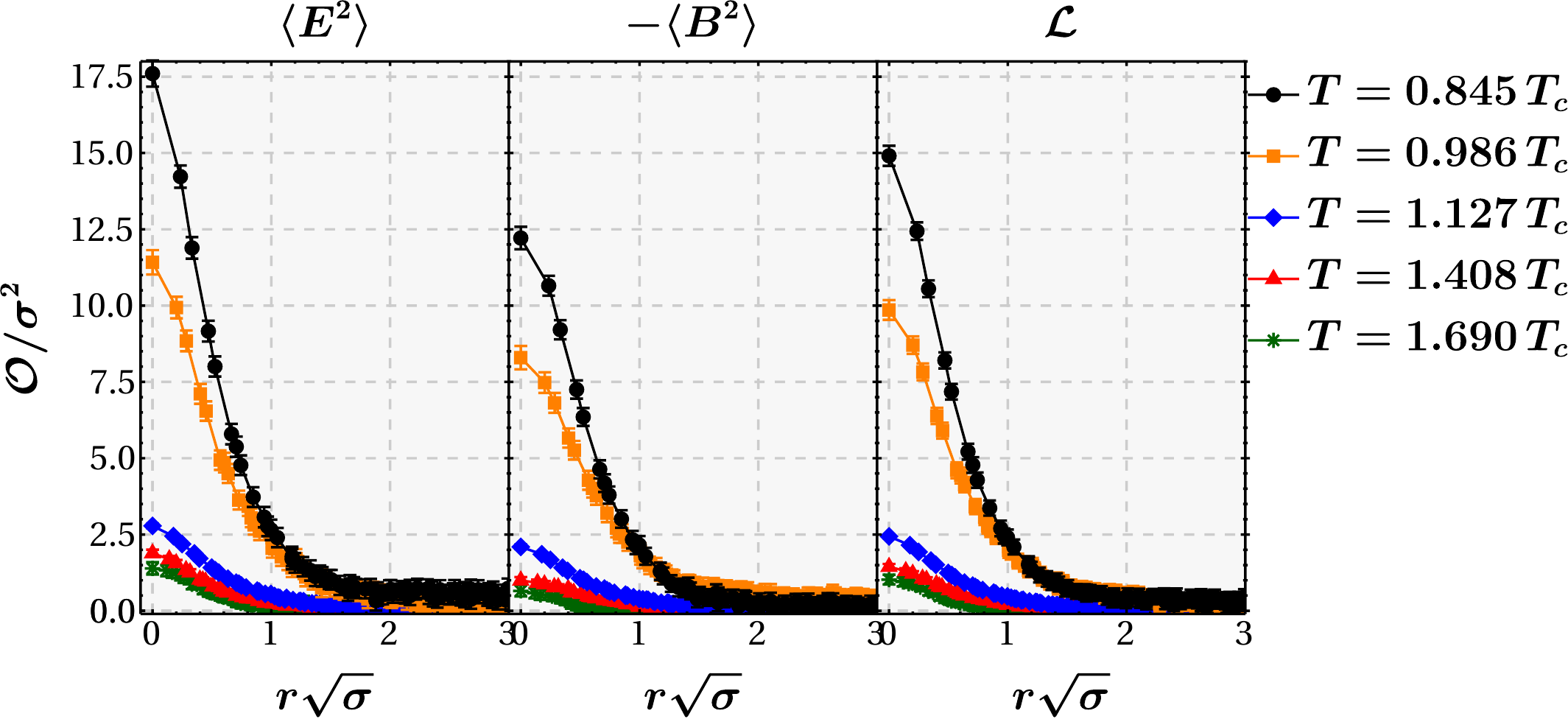}
\par\end{centering}}
\\
\subfloat[$QQ$.\label{fig:FieldSR1Da_1p410_XZ_pp}]{
\begin{centering}
\includegraphics[width=8cm]{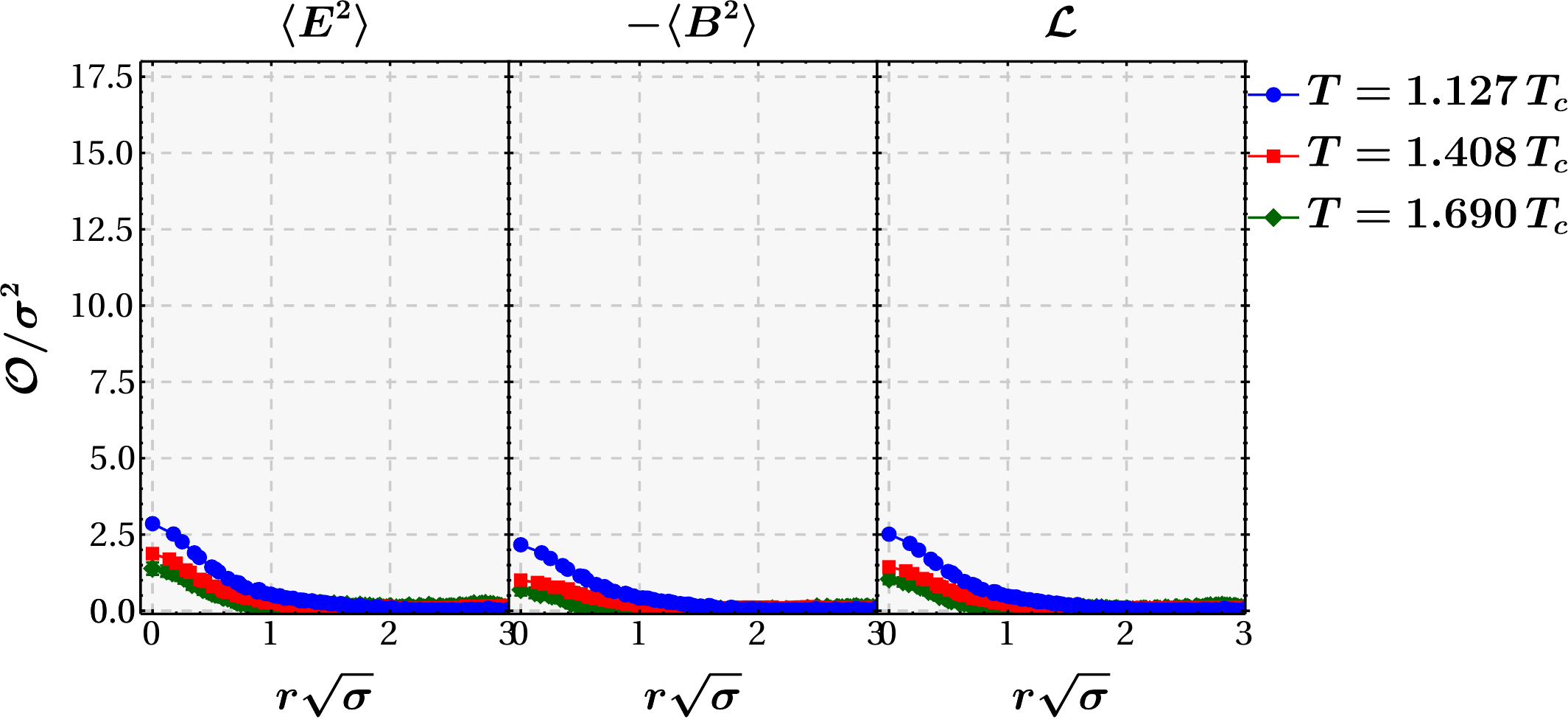}
\par\end{centering}}
\par\end{centering}
\caption{ (Colour Online.) Flux profiles in the mediator plane for $R=1.41\sqrt{\sigma}$, top for the $Q \bar Q$ pair and bottom for a $QQ$ pair. Above the phase transition temperature $T_c$, the $QQ$ and $Q \bar Q$ squared field densities are almost identical modulo error bars.}
\label{fig:profiles_pp_ppdagger}
\end{figure}

In this section, we analyse the flux tube profiles in the mediator plane, equidistant between the charges. Examples of profiles are shown in Fig. \ref{fig:profiles_pp_ppdagger}, where we compare the $Q \bar Q$ and $Q Q$ profiles. 

At $T$ below $T_c$ only the colour singlet $ Q \bar Q$ system produces finite Polyakov loops. Moreover, at $T$  above $T_c$ the profiles \textcolor{black}{seem to be} additive, in the sense the $QQ$ profile is almost identical to the $Q \bar Q$ profile, as discussed in Section III. Thus, in this Section, we specialize in the profiles of the $Q \bar Q$ system only. 

Moreover, we combine the squared field densities in the Lagrangian density to get a clearer signal, with smaller statistical errors. We also make use of the axial discrete symmetry to increase the statistics of points with equal distance $r=\sqrt{x^2 + y^2}$ to the axial charge axis $z$. Our main concern is to compute quantitative results from the profiles for different temperatures $T$ and inter-charge distances $R$.

\subsection{Ansatz for the field density profile}
We first fit the flux tube profile \textcolor{black}{of our squared field density $F^2$} with the ansatz proposed in Ref. \cite{Cardoso:2013lla},
\begin{equation}
\label{eq:ansatz}
F^2(r) = 
{F_0}^2 \exp \left(-{2 \over \lambda} \sqrt{r^2 + \nu^2} +  2 {\nu \over \lambda} \right) + {\cal K} \ ,
\end{equation}
with three physical parameters: the  axis central intensity of the flux tube ${F_0}^2$, the penetration length at large distances from the axis $\lambda$ and the parameter $\nu$ related to the second derivative $- 2 {F_0}^2  / ( \lambda \nu )$. We also have the unphysical parameter $\cal K$ which is due to the statistical fluctuations of the fields at the reference point $\mathbf r_\text{ref}$ of \cref{eq:fmunucomp} 
\textcolor{black}{and due to its finite distance from the centre of the flux tube. The parameter $\cal K$ accounts for the error from the (arbitrary) choice of the reference point, it is small since it is vanishing for high statistics and the profile decreases at least exponentially with $|\mathbf r - \mathbf r_\text{ref}|$. }

Moreover, with our fit we also compute another quantitative parameter \cite{Cardoso:2013lla}, considering the normalized $\left[ F^2(r)- {\cal K} \right]$ as a profile density,
the root mean square width,  $w= \sqrt{ \langle r^2 \rangle }$,  of the flux tube profile,
\begin{eqnarray}
w^2
 & = &
 \int _0^\infty r^3 \, \left[ F^2 (r) - {\cal K} \right] \, dr \over  \int _0^\infty r \, \left[ F^2 (r) - {\cal K} \right]\,  dr
  \nonumber
 \\
& =& 
{3 \over 2} \lambda^2 
+ 2 {  \lambda  \nu^2  \over \lambda + 2  \nu }  
\ .
\label{eq:width}
\end{eqnarray}

\subsection{Computation of the systematic errors}
To compute the width, $w$, which is the main quantitative result of this paper, we first must choose what components $\langle {E_i}^2 \rangle $ and $ \langle {B_i}^2 \rangle $ we adopt as profile density $F^2$. Note all components have a similar behaviour, and thus we can choose their most favourable linear combination. We opt for the Lagrangian density $\cal L$ in \cref{eq:lagrangiand}, which has a better signal-to-noise ratio.

\begin{figure}[!t]
\captionsetup[subfloat]{farskip=0.1pt,captionskip=0.1pt}
\begin{centering}
\includegraphics[width=8cm]{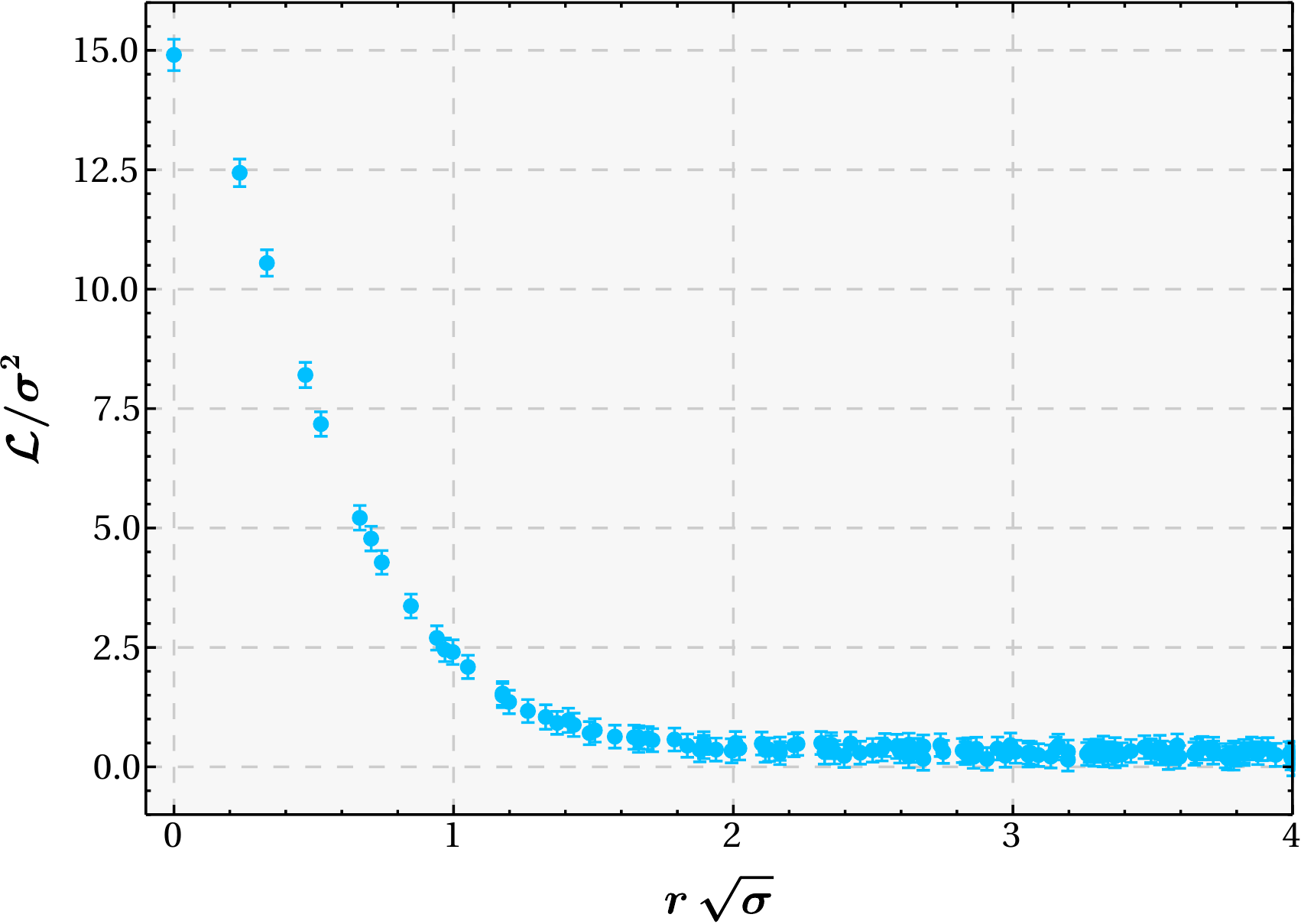}
\includegraphics[width=8cm]{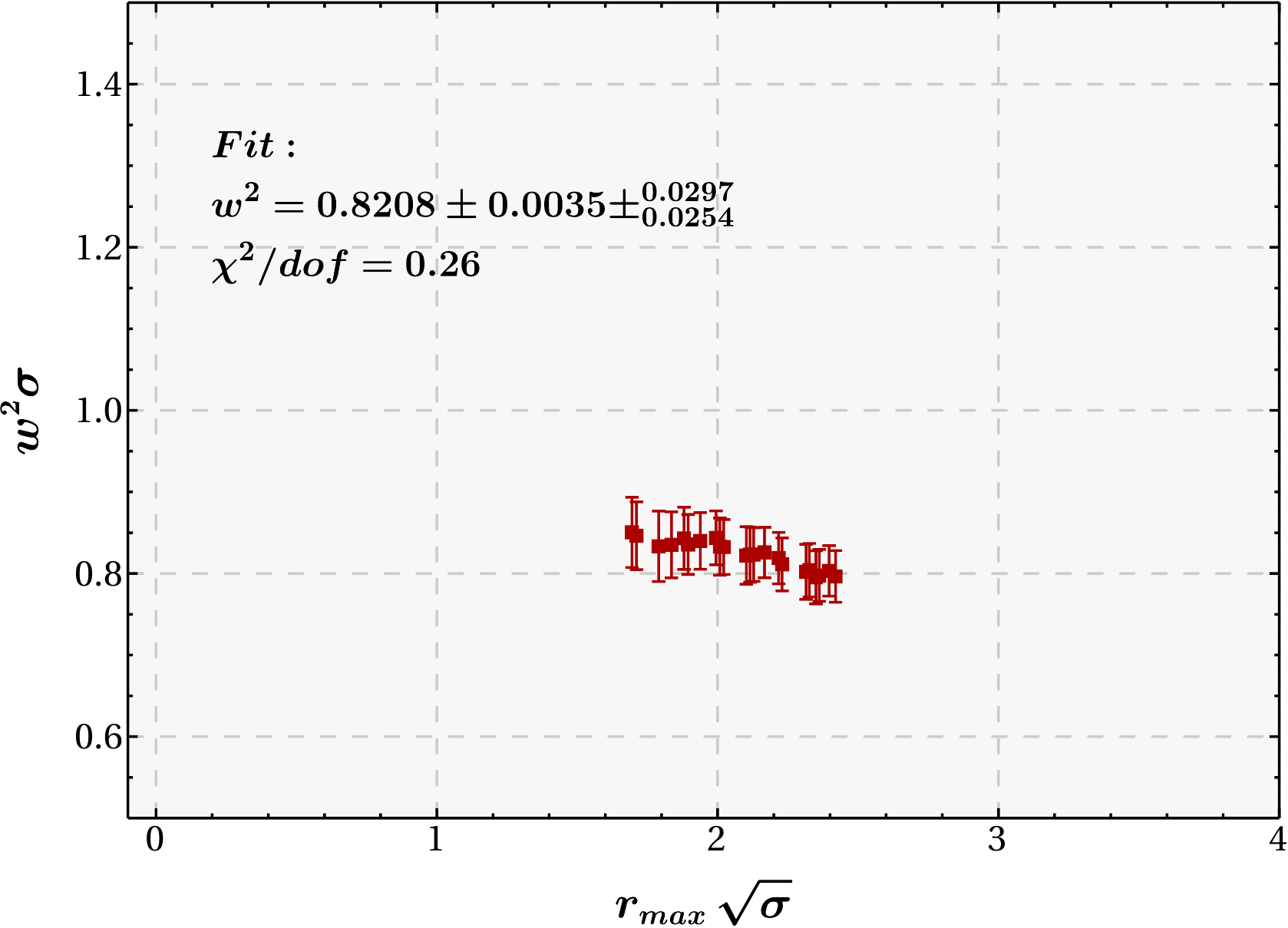}
\par\end{centering}
\caption{(Colour Online.) 
\textcolor{black}{
Systematic error computation, illustrated  for $\beta=5.96$ and $R=6 a$.
In the top panel we show the profile of the flux tube Lagrangian density $\cal L$ as a function of $r$.
In the bottom panel we show the flux tube width $w$ as a function of $r_\text{max}$, with statistical error bars. 
The statistical error bars of the width $w$ are computed with the average, maximum and minimum values of $w$ in the $r_\text{max}$ interval.
To compute the systematic error bars we consider the interval $r_\text{max} \in \left[ r \left({\cal L}_0 /50 \right) , r \left({\cal L}_0 /500 \right) \right] $.}
}
\label{fig:width_fit3}
\end{figure}

\begin{table}[!t]
\begin{ruledtabular}
\begin{tabular}{ccccccc}
\T\B$R$ & $R\sqrt{\sigma}$ & $w^2(R/2) \sigma$ & combined error\\
\T\B 6 & 1.4101 & $0.820834(35) \left(_{254}^{297}\right)$ & 0.0310 \\
\T\B 8 & 1.8802 & $0.889802(18) \left(_{239}^{451}\right)$ & 0.0363 \\
\T\B 10 & 2.3502 & $0.956235(27) \left(_{440}^{1085}\right)$ & 0.0790 \\
\T\B 12 & 2.8203 & $1.16461(19) \left(_{669}^{1455}\right)$ & 0.1081 \\
\end{tabular}
\end{ruledtabular}
\caption{Results for the flux tube width with statistical and systematic errors, in the case of $\beta=5.96$.}
\label{tab:}
\end{table}

\begin{figure*}[!t]
\captionsetup[subfloat]{farskip=0.1pt,captionskip=0.1pt}
\begin{centering}
\subfloat[Central axial ${\cal L}_0$ as a function of inter-charge distance.\label{fig:L0_all}]{
\begin{centering}
\includegraphics[width=8cm]{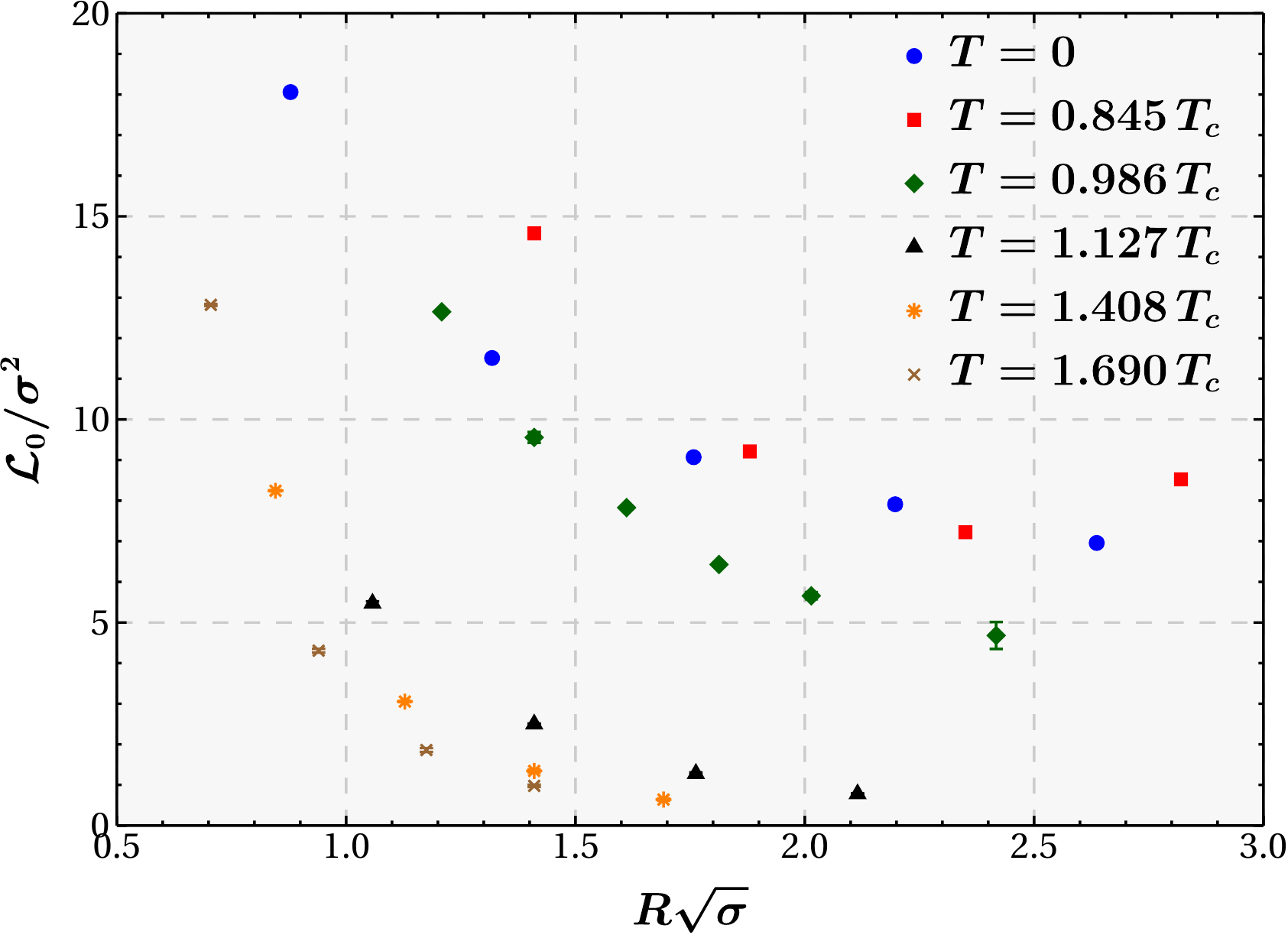}
\par\end{centering}}
\subfloat[Temperature dependence of ${\cal L}_0$ for fixed $R=1.41\sqrt{\sigma}$.\label{fig:L0_1p41}]{
\begin{centering}
\includegraphics[width=8cm]{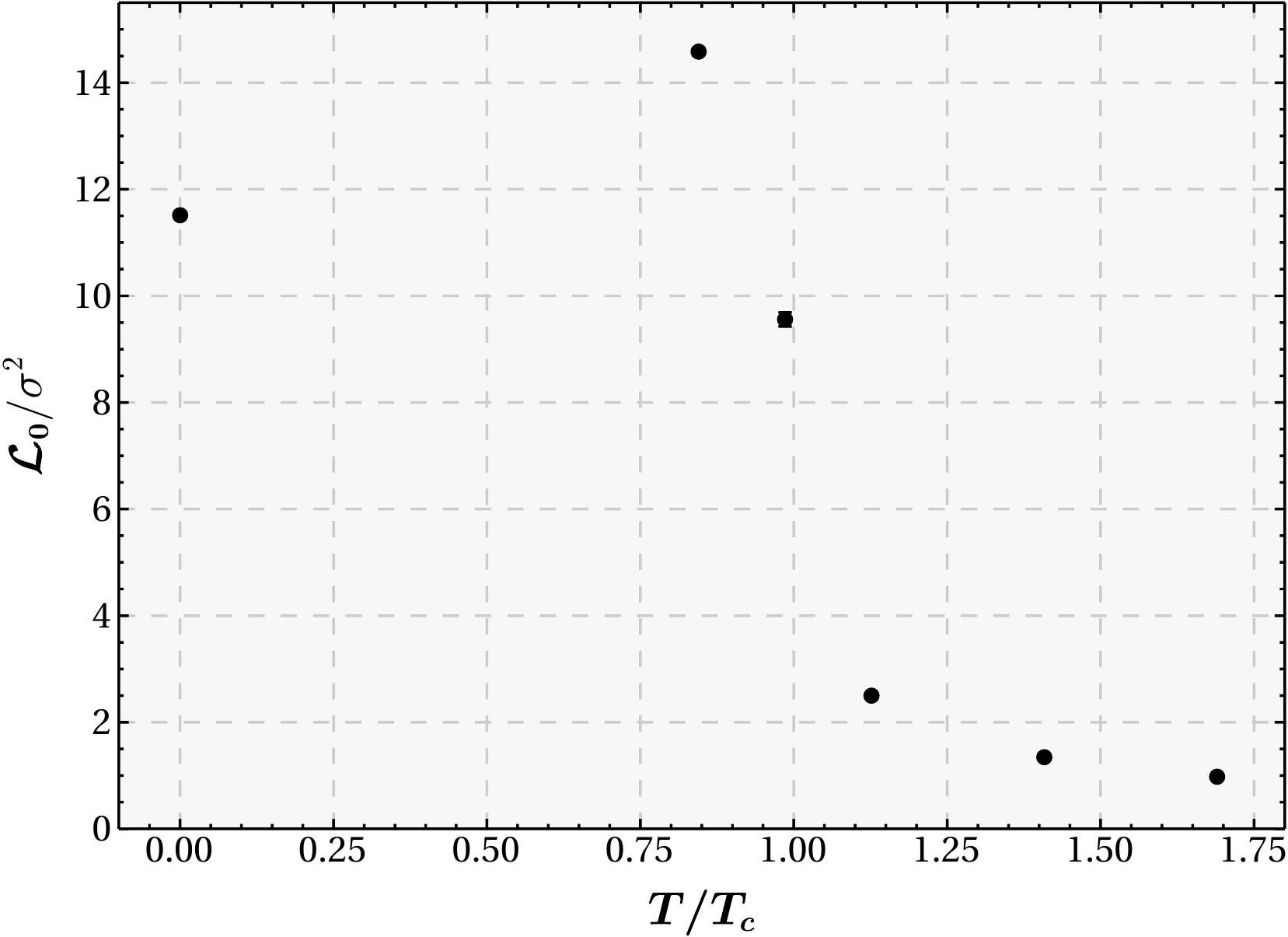}
\par\end{centering}}
\par\end{centering}
\caption{ (Colour Online.) Central value parameter of the Lagrangian density ${\cal L}_0$ as a function of the $Q \bar Q$ inter-charge distance $R$ in string tension units, for all our the different temperatures (left) and as a function of temperature for a fixed distance of $1.41 \sqrt \sigma$ (right). The error bars are total error bars, including both statistical and systematic errors.}
\label{fig:L0}
\end{figure*}

\begin{figure*}[!t]
\captionsetup[subfloat]{farskip=0.1pt,captionskip=0.1pt}
\begin{centering}
\subfloat[Width $w^2$ as a function of inter-charge distance..\label{fig:width_all}]{
\begin{centering}
\includegraphics[width=8cm]{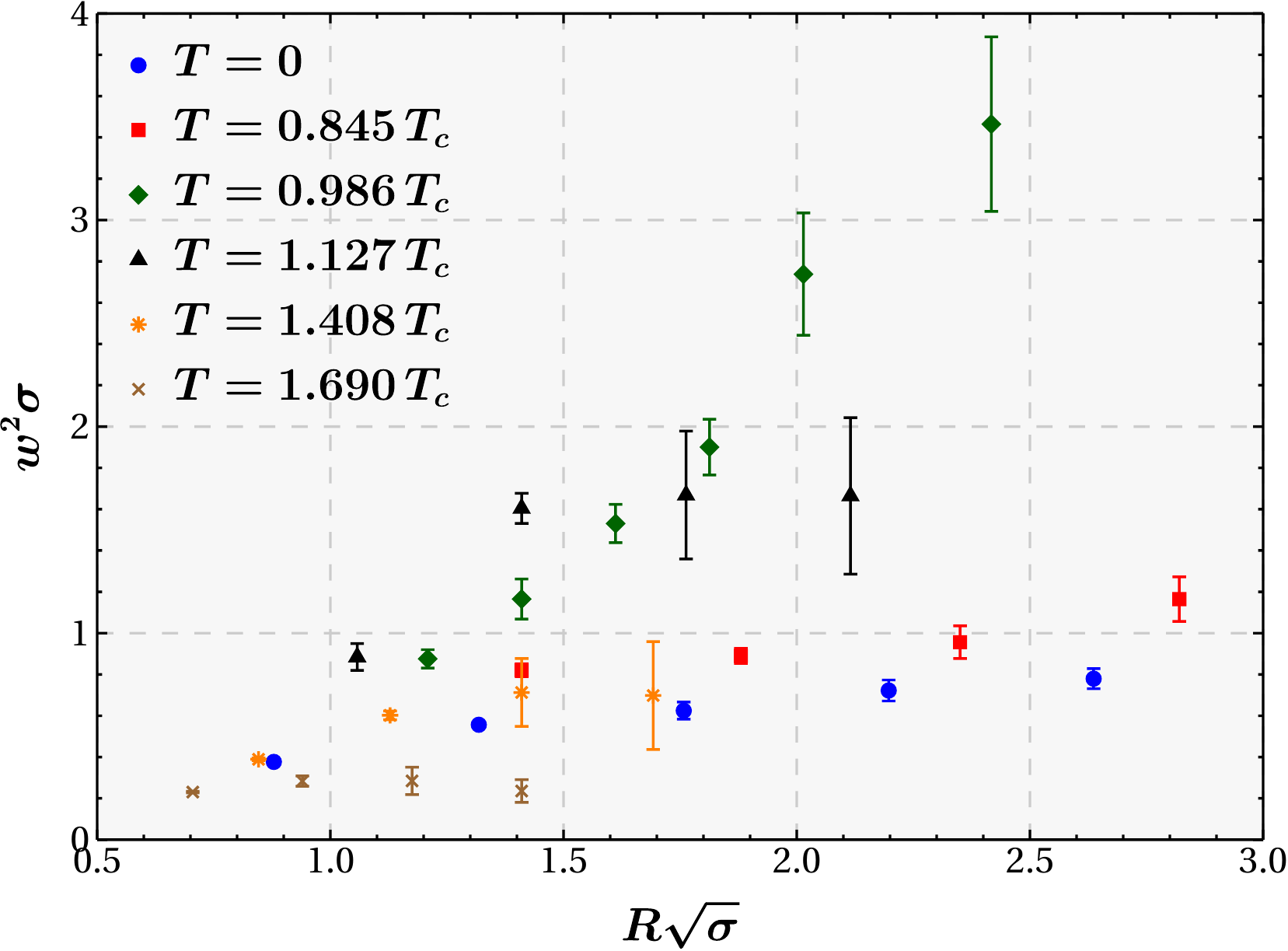}
\par\end{centering}}
\subfloat[Temperature dependence of $w^2$ for fixed $R=1.41\sqrt{\sigma}$.\label{fig:width_1p41}]{
\begin{centering}
\includegraphics[width=8cm]{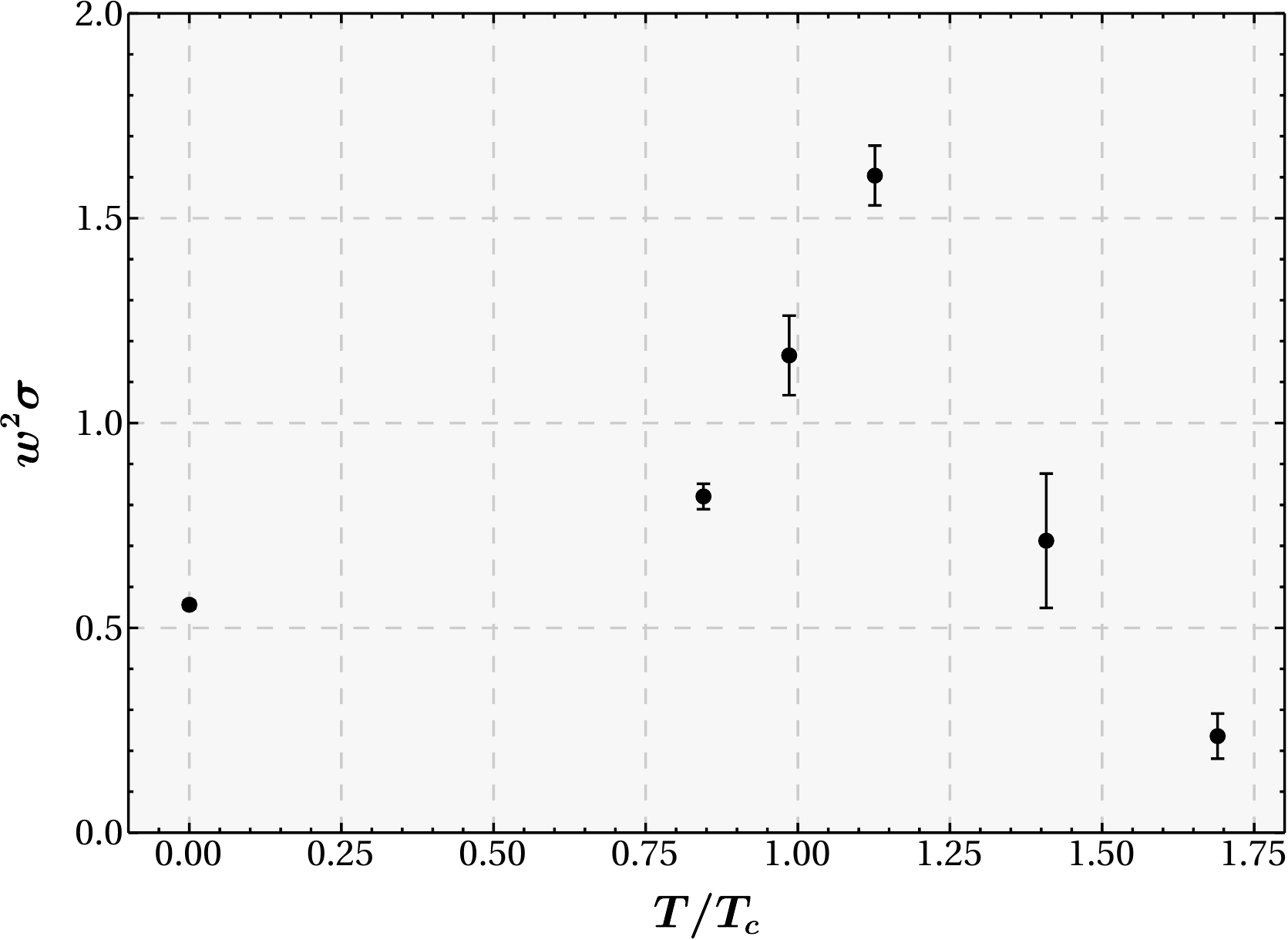}
\par\end{centering}}
\par\end{centering}
\caption{ (Colour Online.) Flux tube square width $w^2$ as a function of the $Q \bar Q$ inter-charge distance $R$ in string tension units, for all our the different temperatures (left) and as a function of temperature for a fixed distance of $1.41 \sqrt \sigma$ (right). The error bars are total error bars, including both statistical and systematic errors.}
\label{fig:width}
\end{figure*}

\textcolor{black}{
Moreover there are systematic errors present in our fit of $F^2(r)$ and in the computation of the width $w$.
}
We must select the interval $\left[ 0, r_\text{max} \right]$ in the distance $r$ where we fit the profile of the flux tube. This leads to a systematic error, as in Ref. 
\cite{Cichy:2012vg,Bicudo:2015vta,Bicudo:2015kna}.
Because the profile vanishes at least exponentially with distance, the points with larger $r$ correspond to a vanishing profile, with no relevant physical information. Moreover, the statistical error increases with distance. We estimate the most significant intervals are the ones where $r_\text{max}$ corresponds to a value for the Lagrangian density between ${\cal L}_0 / 50$ and ${\cal L}_0 / 500$. For instance, in the case of a Gaussian distribution, this would correspond to a \textcolor{black}{fraction of distribution included} between 99.5 \% and 99.96 \%.

To extract the width from the Lagrangian density, as illustrated in  \cref{fig:width_fit3}, we proceed as follows. We crudely estimate ${\cal L}_0$ from  the point in the charge axis, at $r=0$. We fit the Lagrangian density with our ansatz in Eq. (\ref{eq:ansatz}), with $r_\text{max}$ in the interval between ${\cal L}_0 / 50$ and ${\cal L}_0 / 500$ crudely estimated. With our fit we then get a correct estimate of  ${\cal L}_0, \ {\cal L}_0 / 50$ and ${\cal L}_0 / 500$, and redo our fits with the correct $r_\text{max}$. Then, for all possible different intervals with $r_\text{max} \in \left[ {\cal L}_0 / 50 , {\cal L}_0 / 500 \right]$, and respective fits of $\cal L$, we determine the parameters ${F_0}^2={\cal L}_0 , \ \lambda, \ \nu$ and $w^2$.

\begin{figure*}[!t]
\captionsetup[subfloat]{farskip=0.1pt,captionskip=0.1pt}
\begin{centering}
\subfloat[$T=0.845\,T_c$.\label{fig:width_fit_5p96}]{
\begin{centering}
\includegraphics[width=8cm]{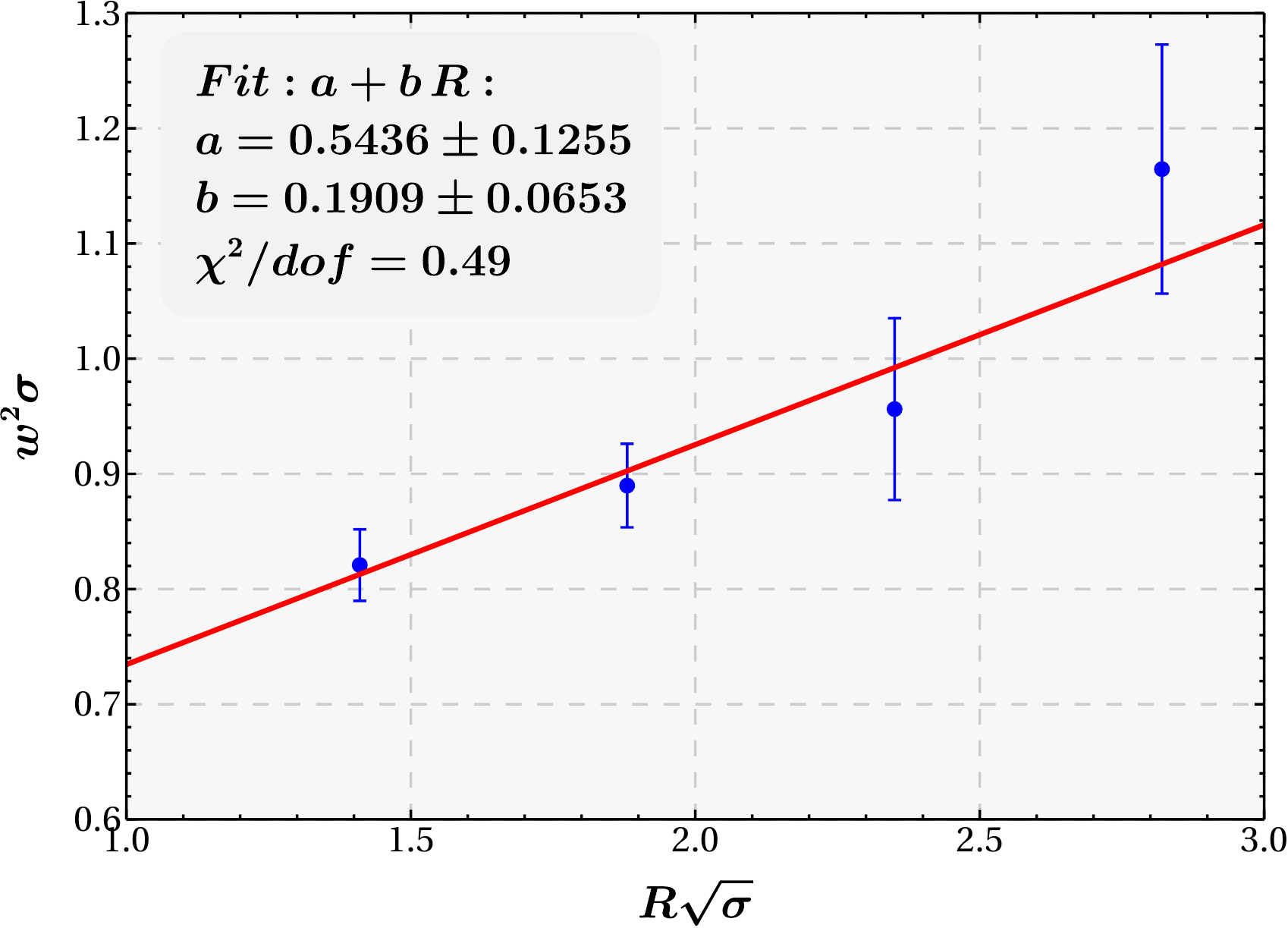}
\par\end{centering}}
\subfloat[$T=0.986\,T_c$.\label{fig:width_fit_6p0534}]{
\begin{centering}
\includegraphics[width=8cm]{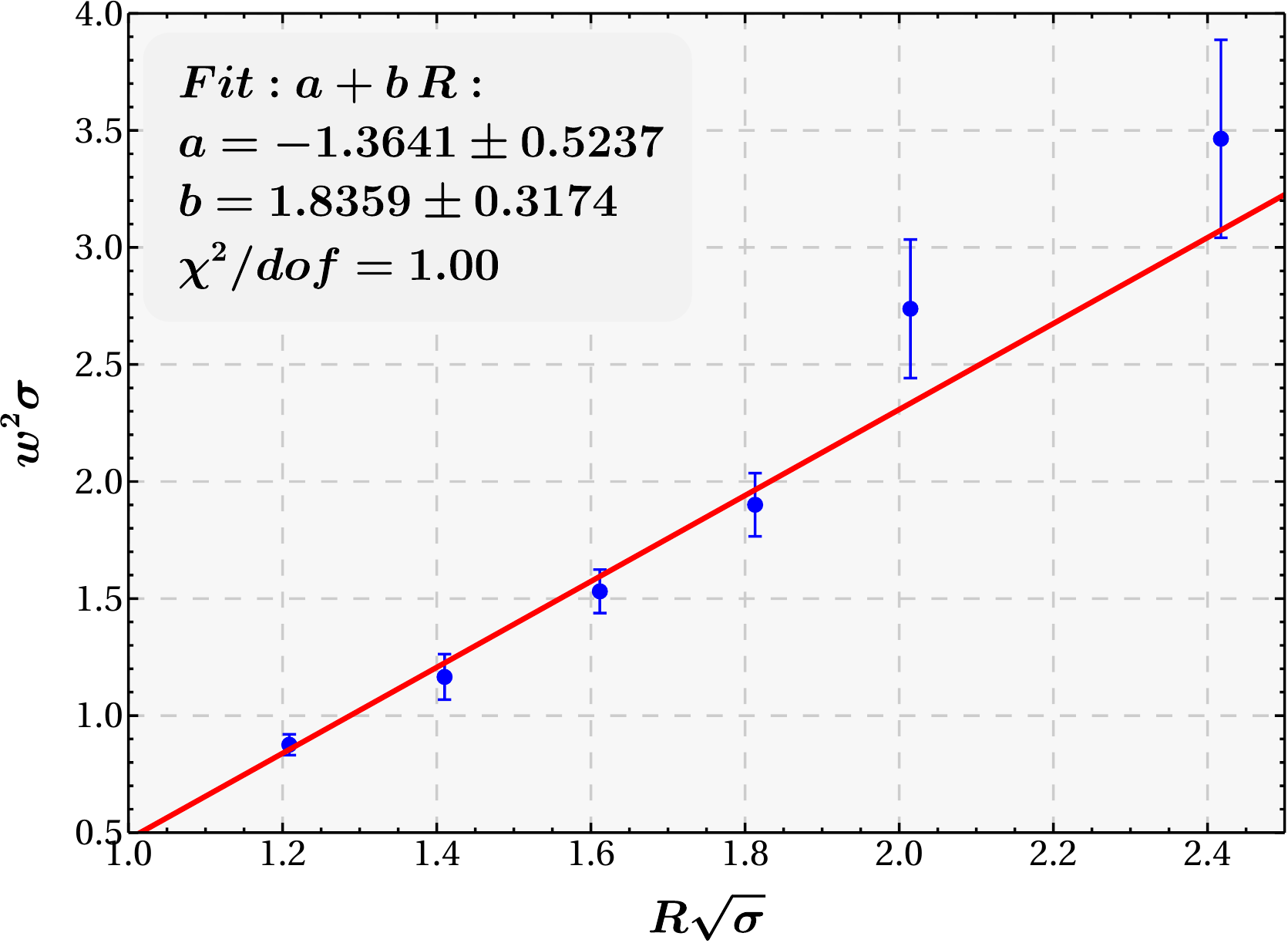}
\par\end{centering}}
\par\end{centering}
\caption{  (Colour Online.) Results with combined statistical and systematic errors, at finite temperature below $T_c$, for the widening of the flux tube as a function of the inter-charge $Q \bar Q$ distance.}
\label{fig:width_fit_Tfin}
\end{figure*}

\begin{figure}[!t]
\captionsetup[subfloat]{farskip=0.1pt,captionskip=0.1pt}
\begin{centering}
\includegraphics[width=8cm]{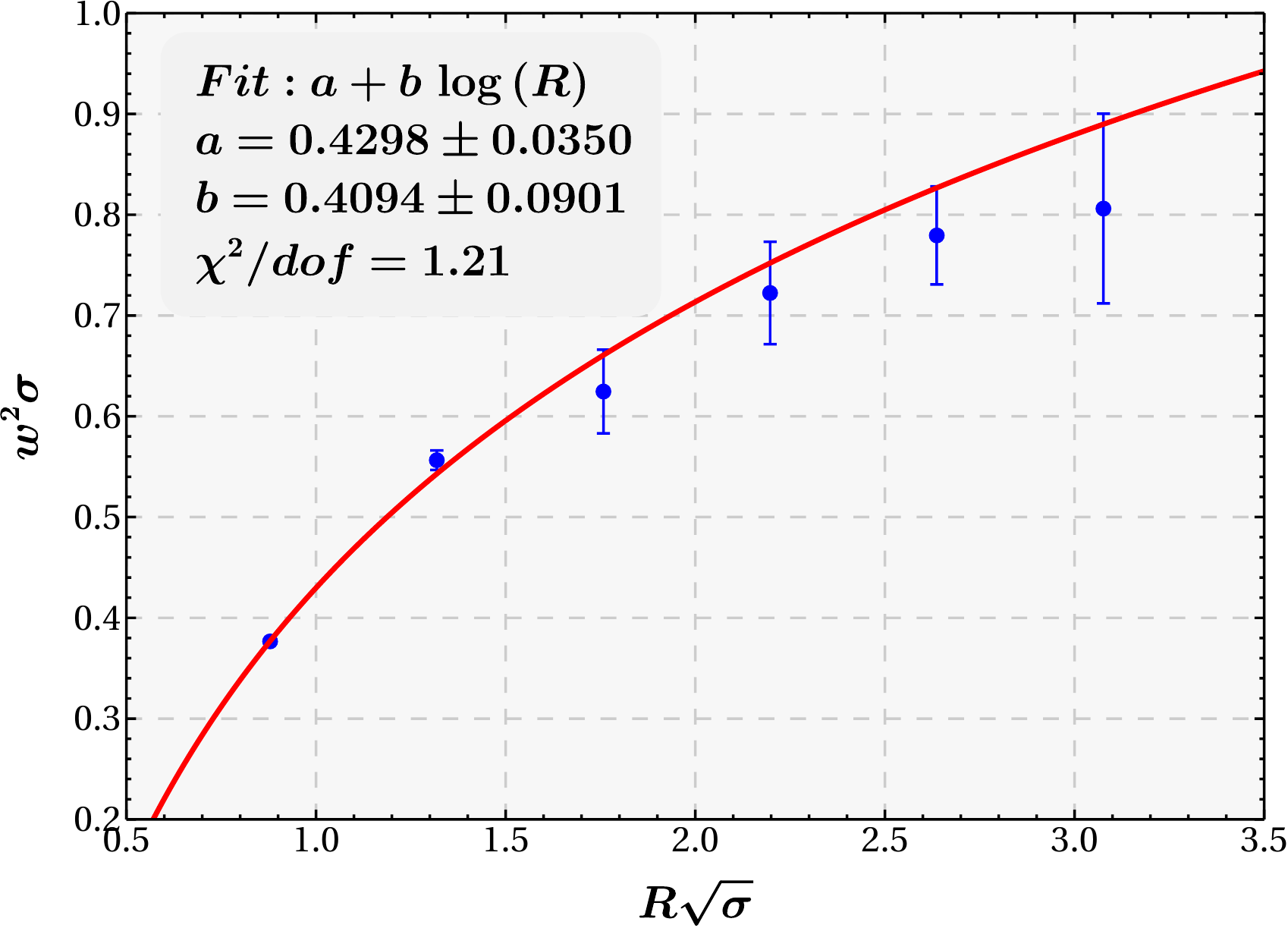}
\par\end{centering}
\caption{ (Colour Online.) Results with combined statistical and systematic errors, at vanishing $T=0$, for the widening of the flux tube as a function of the inter-charge $Q \bar Q$ distance.}
\label{fig:width_fit_T0}
\end{figure}

The parameters come with a statistical error from each fit range $\left[ 0, r_\text{max} \right]$. Moreover, combining all fit ranges, we obtain a systematic error  
\cite{Cichy:2012vg,Bicudo:2015vta,Bicudo:2015kna}. We obtain a barycentre for the systematic error value, and two different upper and lower error bars considering the maximum differences to the barycentre. Finally, for the total error bars, the systematic upper and lower error are averaged, and combined with the statistical error to provide a total error. In \cref{tab:} we detail the statistical, systematic and total error bars in the case of $\beta=5.96$.

\subsection{Results for the width and central density}

{\color{black}
We find the central value parameter ${\cal L}_0$ and the width $w$ have small statistical and systematic error bars.
However, our data is not precise enough to determine both the parameters $\lambda$ and $\nu$ with small error bars. In some of our cases, these two parameters have large error bars, due to redundancy. To remove this redundancy, we would need more data, in order to reduce the statistical error bars.

We show all our results for the  central value parameter ${\cal L}_0$ in \cref{fig:L0}. It is clear the central value of the flux tube density ${\cal L}_0$ has a large step downwards when $T_c$ is crossed. A gap is clearly visible between the $T< T_c$ and the $T>T_c$ data in \cref{fig:L0}. 

Moreover we show all our results, for all distances and temperatures, for the  squared width $w^2$ in \cref{fig:width}.
While at temperatures below $T_c$ the width is clearly growing with distance, there is apparently no evidence for widening at temperatures above $T_c$ since the square widths are apparently constant. 

Finally, we show a detailed analysis of the widening of the flux tubes, expected to occur only at temperatures $T<T_c$. We plot the square width $w^2$, in three separated plots for our three temperatures $T<T_c$, as a function of the $Q \bar Q$ inter-charge distance $R$, in \cref{fig:width_fit_T0,fig:width_fit_Tfin}.

 For the finite $0 < T < T_c$ data computed here, we observe in \cref{fig:width_fit_Tfin} for the first time a $SU(3)$ behaviour previously studied for instance in compact U(1) \cite{Amado:2012wt}. Indeed our data is consistent with a linear fit, as predicted by Ref.   \cite{Allais:2008bk}. 

We also re-analyse in \cref{fig:width_fit_T0} the data of Ref.  \cite{Cardoso:2013lla} for $T=0$ with the present technique to compute the systematic errors, and we confirm the logarithmic behaviour of the width. 
}

\section{ Conclusions}

We compute the square densities of the chromomagnetic and chromoelectric fields produced by different Polyakov loop sources, above and below the phase transition. We fit the profile of the $Q \bar Q$ flux tubes and compute physical parameters, including the flux tube width, with statistical and systematic errors.

As the distance increase between the sources, the fields square densities decrease.
Below the deconfinement critical temperature, this decrease is moderate and is consistent with the widening of the flux tube  as already seen in studies at zero temperature \cite{Cardoso:2013lla}, moreover the field intensity clearly decreases when the temperature increases, as expected from the critical curve for the string tension \cite{Cardoso:2011hh}.

Above the deconfinement critical temperature, at $T>T_c$, the fields rapidly decrease to zero as the quarks are pulled apart, qualitatively consistent with screened Coulomb-like  fields.
While the width of the flux tube below the phase transition temperature increases with the separation between the quark-antiquark, above the phase transition we find no evidence for widening.
Moreover, the squared field densities are additive, in the sense the fields produced by a quark $Q$, an antiquark $\bar Q$ and a colour adjoint source $A$ approximately add up together when these sources coexist. In the same perspective, the $QQ$ and the $Q \bar Q$ square fields are essentially similar. This is in contradiction with the squeezing of the fields into a flux tube which should be non-linear. Thus we find evidence for the non-existence of flux tubes at temperature above the deconfinement temperature, $T>T_c$.

As an outlook, it would be interesting to complete the present study with further tests of the additive nature of squared field densities above $T_c$. We also would like to produce results with smaller error bars, in order to be able to measure precisely the penetration length parameter $\lambda$ at finite $T$, as we did for vanishing $T$ in Ref. \cite{Cardoso:2013lla}.  We also plan to produce the different Polyakov loop - Polyakov loop potentials, relevant for modelling the deconfinement transition
\cite{Smith:2013msa,Bicudo:2014cra,Greensite:2014isa}.
\textcolor{black}{It would also be interesting  to observe the cross-over between a logarithmic widening at small T and a linear widening at larger $T<T_c$ as in Eq. (\ref{eq:linearfiniteT}). It will be necessary to develop a new technique 
\cite{Koma:2017hcm}
to match the $T=0$ Wilson loops \cite{Cardoso:2013lla} with the higher $T$ Polyakov loops computed here. }

\acknowledgments
Nuno Cardoso and Marco Cardoso are supported by FCT under the contracts SFRH/BPD/109443/2015 and SFRH/BPD/73140/2010 respectively.
We also acknowledge the use of CPU and GPU servers of PtQCD, supported by NVIDIA, CFTP and FCT grant UID/FIS/00777/2013.
Our computations are performed in NVIDIA GPUs only, using the CUDA language.


\bibliographystyle{apsrev4-1}
\bibliography{fluxtube_T}{}

\end{document}